\documentclass[preprint,12pt]{elsarticle}

\usepackage{graphicx}

\usepackage{amssymb}
\usepackage{amsthm}
\usepackage{amsmath}
\usepackage{amsfonts}
\usepackage{bm}

\usepackage{natbib}        
\usepackage{epstopdf}
\usepackage{epsfig}
\usepackage{comment}

\usepackage[nodots]{numcompress}

\journal{Computers and Operations Research}

\DeclareMathOperator*{\argmax}{arg\,max}	

\newtheorem*{definition}{Definition}

\begin{document}

\begin{frontmatter}



\title{Finding Optimal Strategies in a Multi-Period Multi-Leader-Follower Stackelberg Game Using an Evolutionary Algorithm}


\author[ad1]{Ankur Sinha\corref{cor1}} \cortext[cor1]{Corresponding author}
\ead{ankur.sinha@aalto.fi}

\author[ad1]{Pekka Malo}
\ead{pekka.malo@aalto.fi}

\author[ad1]{Anton Frantsev}
\ead{anton.frantsev@aalto.fi}

\author[ad2]{Kalyanmoy Deb}
\ead{kdeb@egr.msu.edu}

\address[ad1]{Department of Information and Service Economy, Aalto University School of Business,\\ Helsinki, P.O. Box 21210,
FI-00076 AALTO, Finland}
\address[ad2]{Department of Electrical and Computer Engineering, Michigan State University,\\ East Lansing,
MI-48824, USA}

\begin{abstract}
Stackelberg games are a classic example of bilevel optimization problems, which are often encountered in game theory and economics. These are complex problems with a hierarchical structure, where one optimization task is nested within the other. Despite a number of studies on handling bilevel optimization problems, these problems still remain a challenging territory, and existing methodologies are able to handle only simple problems with few variables under assumptions of continuity and differentiability. In this paper, we consider a special case of a multi-period multi-leader-follower Stackelberg competition model with non-linear cost and demand functions and discrete production variables. The model has potential applications, for instance in aircraft manufacturing industry, which is an oligopoly where a few giant firms enjoy a tremendous commitment power over the other smaller players. We solve cases with different number of leaders and followers, and show how the entrance or exit of a player affects the profits of the other players. In the presence of various model complexities, we use a computationally intensive nested evolutionary strategy to find an optimal solution for the model. The strategy is evaluated on a test-suite of bilevel problems, and it has been shown that the method is successful in handling difficult bilevel problems.

\end{abstract}


\end{frontmatter}



\section{Introduction}
\label{sec:intro}
The relevance of bilevel optimization problems has been amply recognized by researchers and practitioners alike \cite{bilevel-book}. The key difference between bilevel programming problems and other optimization problems is their nested structure. A bilevel program is commonly defined as an optimization problem, which contains another optimization task within the constraints of the outer problem. The outer optimization problem is generally termed as the upper level problem, and the constraining optimization task is denominated as the lower level problem. Such a nested structure means that a solution to the upper level problem is considered feasible only if it is an optimal solution to the lower level problem. Because of this requirement, bilevel optimization problems can quickly become very difficult to solve.

Bilevel optimization problems often appear as leader-follower problems in the fields of game theory and economics. When formulating a leader-follower scenario as a bilevel programming problem, the leader's optimization task is modeled at the upper level, constrained by the follower's optimization task at the lower level. The leader has the ability to move first, and is assumed to possess all necessary information about the follower's possible reactions to the actions taken by the leader. The follower, on the other hand, observes the leader's actions before reacting optimally to them. By solving a Stackelberg competition model, the leader can forecast the follower's reactions and determine his own optimal actions.

In this paper, we model a particular kind of oligopolistic market involving multiple leaders and followers interacting over multiple time periods. Such a scenario can be analysed as a special case of a multi-period multi-leader-follower Stackelberg game. This problem builds on the original definition of duopolistic competition proposed by \citet{Stackelberg52} and its various extensions \cite{Sherali84,Sherali83,cao12,DeMiguel09}. The multi-period market model presented in this paper allows us to incorporate relevant dynamics and several practical aspects, such as investment and marketing effects, in a flexible manner. As an extension to the prior work, we also assume the production of all players in this market to be discrete variables while allowing the investment and marketing decisions to be continuous. This feature makes the model framework applicable to markets where the products are valuable and are not well approximated by continuous variables, such as in the aircraft manufacturing market. Further realism is added to the problem in the form of non-linear demand and cost functions, and constraints on investment and marketing budget for both the leaders and the followers. The effects of investment and marketing are considered to be cumulative leading to interactions between the time-periods. Evolutionary algorithms have been used in the past in the domain of game theory with players making simultaneous moves (Nash Equilibria), players making asynchronous moves, or a mix of the two situations. For instance, Lung et al. \cite{lung08} consider a multi-player game and propose a domination concept to detect the Nash-equilibria using an evolutionary multi-objective algorithm. Koh \cite{koh12} proposes an evolutionary algorithm for handling a similar model with multiple leaders and followers as considered in this paper. However, Koh assumes that it is possible to write the first-order conditions for the lower level problem, which is not possible in the problem considered in this paper. Classical techniques like branch-and-bound have also been applied to handle bilevel problems involving multiple players. Lu et al. \cite{lu07} use a branch-and-bound technique to handle linear Stackelberg problems with a single leader and multiple followers. Zhang et al. \cite{zhang08} study a similar problem with a single leader and multiple followers, but in this study the functions involve fuzzy coefficients and multiple objectives at both levels.

Many studies have been conducted in the field of bilevel programming \cite{colson,vicente-review,dempe-dutta,my-ecj10} and on its practical applications for solving various problems \cite{bilevel-book}. Aspects of bilevel programming have also been reviewed in detail by \citet{colson} and \citet{vicente-review}. The problems encountered in the literature are commonly solved with the help of approximate solution methodologies \cite{bianco-kkt,Dempe02,stackelberg-design}. These techniques work with numerous simplifying assumptions, and most of them are inapplicable to problems with higher levels of complexity. The classical methods widely employed by researchers and practitioners to solve these problems include the Karush-Kuhn-Tucker approach \cite{bianco-kkt,bilevel-KKT1}, Branch-and-bound techniques \cite{bard82} and the use of penalty functions \cite{aiyoshi81}. However, the recent technological advances and increased computing power have made heuristic approaches, such as evolutionary algorithms, more popular for solving complex optimization problems. Evolutionary algorithms have also been applied to bilevel programming problems \cite{cao12,yin-bilevel,GA_Wang,my-ifac09,my-ecj10}. Given the encouraging results obtained with such techniques, we have also opted to employ a nested evolutionary algorithm to solve the multi-leader-flower problem. The choice is well motivated by the complexities due to non-linearity and discreteness that are inherent in our model for oligopolistic markets.

The remainder of the paper is structured as follows. Section \ref{sec:model} presents a general formulation of a multi-period multi-leader-follower Stackelberg competition model. Section \ref{sec:multiplayer} outlines a more concrete model with a discussion about its possible application to the aircraft manufacturing market. Section \ref{sec:method} contains a description of the procedure followed by the nested bilevel evolutionary algorithm to arrive at the optimal solution, and then in Section \ref{sec:evaluation} we evaluate the proposed nested bilevel evolutionary algorithm. Section \ref{sec:results} presents the results on the multi-period multi-leader-follower Stackelberg optimization problem. Section \ref{sec:convergence} provides a convergence analysis on a simplistic version of the Stackelberg competition model. Lastly, the conclusions and plans for future work are summarized in Section \ref{sec:conclusion}.

\section{Generalized Stackelberg Competition Model}
\label{sec:model}

Three main models -- the Bertrand, Cournot, and Stackelberg competition models -- are extensively applied in economics when modeling multi-firm competition \cite{Sherali83,DeMiguel09,Sawaya11}. The first two models are generally used when the competing firms have an approximately equal amount of market power, and make production and pricing decisions simultaneously. However, in an oligopoly, where some of the competing firms have more market power than others, a Stackelberg model is considered more suitable. Such markets are much more common in practice than strict oligopolies and deserve more attention \cite{Sherali84,DeMiguel09}. In this section, we outline some of the notations used throughout the rest of this paper and present a generalized formulation of a Stackelberg competition model with multiple leaders and followers.


In the context of the multi-leader-follower problem framework, a strategy is defined as a sequence of decisions made by a player throughout the duration of the game. A strategy for a leader $i$ can be stated as $s_{l,i}~=~\bigl(s_{l,i}^t \bigr)_{t=1}^{T} \in S_{l,i}$, where $s_{l,i}^t$ is a particular decision made by the leader at time $t$, and $S_{l,i}$ is the set of all alternative strategies available to leader $i$. 
Similarly, we denote a strategy for a follower $j$ as $s_{f,j} = \bigl(s_{f,j}^t \bigr)_{t=1}^{T} \in S_{f,j}$, where $s_{f,j}^t$ is a particular decision of follower $j$ at time $t$, and $S_{f,j}$ is the set of alternative strategies. 
A combination of strategies adopted by $N$ leaders for the entire duration of the game can be presented as $s_l~=~\bigl(s_{l,1}, \dotsc, s_{l,N}  \bigr) \in S_l$, and for $M$ followers as $s_f~=~\bigl(s_{f,1}, \dotsc, s_{f,M}  \bigr) \in S_f$, where the decision spaces $S_l, \, S_f$ are defined as
\[
S_l = \prod_{i=1}^{N} S_{l,i} \quad \mbox{and} \quad S_f = \prod_{j=1}^{M} S_{f,j}.
\]
By optimizing the objectives of the competing firms in these Stackelberg games, we obtain the optimal strategies for the leaders and followers as the solution. 


\begin{definition}[General Stackelberg Competition Model] A general multi-leader-follower Stackelberg competition model with $N$ leaders and $M$ followers may be formulated as
\begin{align}
	\max_{s_l,s_f} \quad & \sum_{i=1}^{N} \Psi_{l,i} (s_l, s_f) \\
	\mbox{s.t.} \quad & s_f = \argmax_{s_f} 
		\left\lbrace \sum_{j=1}^{M} \Psi_{f,j} (s_l,s_f): g_{f,h}(s_l,s_f) \leq 0, h \in \{1,\ldots,H\} \right\rbrace ,\\
	& g_{l,k} (s_l,s_f) \leq 0, k \in \{1,\ldots,K\} ,\\
	& s_l \in S_l, \quad s_f \in S_f ,
\end{align}
where $\Psi_{l,i}, \Psi_{f,j}$ denote the objective functions for leader $i$ and follower $j$ respectively. Similarly, the constraints for the leaders and followers are given by mappings $g_{l,k}$ and $g_{f,h}$. The number of constraints for the leaders and followers is $K$ and $H$ respectively.
\end{definition}
Each of the $N$ leaders acts as a traditional Stackelberg leader towards the follower firms, but as a Cournot firm with respect to the other $N-1$ leaders. The $N$ leaders choose their strategies simultaneously and non-cooperatively among themselves, as in a classic Cournot competition model, but collectively they are in a Stackelberg competition with the follower firms. Similarly, the $M$ followers act as Stackelberg followers towards the leaders and as Cournot firms with respect to each other. Thus, the model consists of two Cournot competitions encompassed by a Stackelberg competition model. The generality of the formulation allows us to easily extend the problem to include as many leaders and followers as necessary, and to make them as different from one another as needed to model a real-world market situation. However, in this paper we assume that the $N$ leaders and $M$ followers are symmetric among themselves.

In the next section, we present a more concrete example of this general model. Additional complexity is introduced into this extended model by incorporating investment and marketing variables into each decision, expenditure constraints, and discrete production variables into the model for both the leaders and the followers.

\section{Multi-Leader-Follower Stackelberg Model}
\label{sec:multiplayer}

A classic Stackelberg game involves a single leader and a single follower, and the game is played for just one period. Such strict duopolies are very difficult, if not impossible, to find in practice. Oligopolies, i.e. markets with a small number of competing firms, are more common. When the markets have more than one distinct leader and follower, the situation corresponds to a multi-leader-follower Stackelberg game. In such settings, the firms with the first movers' advantage become the Stackelberg leaders, and the others act as followers. As already mentioned, in the model considered in this paper, we assume that the group of firms which act as leaders or followers compete in a Cournot fashion within their respective groups.

An example of such a market is the aircraft manufacturing industry. Due to the high market entry costs, the player base is relatively small and stable. The market is dominated by a couple of large players, namely Boeing and Airbus, who satisfy most of the commercial demand for airplanes, and can be safely regarded as the market leaders. Contracts that are small for these market giants are filled by the smaller players, who do not experience the same economies of scale as the leaders and rightfully assume the positions of the followers in this market. Due to the high costs of the products, each firm in the market tries to optimize its decisions to reduce costs and streamline its processes. Furthermore, the players' production quantities are discrete and comparatively small, meaning that the production of one additional unit may lead to significant shifts in profits. While we do not specifically try to model this particular market, the framework presented in this section could be applied to it without much alteration.

There are several aspects that differentiate the model discussed in this section from a traditional Stackelberg game. The main differences are the presence of multiple leaders and followers, the addition of multiple time periods and the interrelation between those periods caused by investment and marketing expenditures made in each period. Additional realism is introduced into the model by requiring the production amounts to be discrete for both the leaders and the followers. This allows us to replicate the behaviour of those markets where unit cost of production is high. It is also worth noting that a player's ability to invest is not equivalent to their ability to borrow or lend extra funds, which implies that the players have to finance their investment and marketing projects from their own revenues. As a further modification, we relax the strict product homogeneity assumption inherent in the original Stackelberg competition model. By assuming that the products are not perfectly homogeneous but still substitutable, we can allow each firm to influence the price of its own goods to a certain degree by means of marketing. 

In the following, we define the variables used in this model for leader $i$ and follower $j$ respectively:
\vspace{-0.5mm}
\begin{itemize} \itemsep1pt \parskip0pt \parsep0pt
	\item $Q_{l,i}$ and $Q_{f,j}$ are the demands for products
	\item $P_{l,i}$ and $P_{f,j}$ are the prices for products
	\item $\Psi_{l,i}$ and $\Psi_{f,j}$ are the profits
	\item $C_{l,i}$ and $C_{f,j}$ are the production costs
	\item $I_{l,i}$ and $I_{f,j}$ are the investment amounts
	\item $M_{l,i}$ and $M_{f,j}$ are the marketing expenditures
	\item $q_{l,i}$ and $q_{f,j}$ are the production levels
\end{itemize}
\vspace{1mm}
A superscript $t$, when added to any of these variables, denotes the value for the variable at time period $t$. A leader's decision in any time period $t$ is made up of choosing the production level, investment amount, and marketing expenditure for that period and can be written as $s_{l,i}^t~=~\bigl(q_{l,i}^t,I_{l,i}^t,M_{l,i}^t \bigr)$. Similarly, a follower's decision in the same time period is made up of his own production level, investment and marketing expenditure and can be written as $s_{f,j}^t~=~\bigl(q_{f,j}^t,I_{f,j}^t,M_{f,j}^t \bigr)$. The production variables of both players are assumed to be discrete, while their investment and marketing expenditures are expressed in monetary terms and are hence continuous. Therefore, the decision spaces of the leaders and followers for any time period $t$ are given by $S_{l,i}^t = \mathbb{K} \times \mathbb{R^+} \times \mathbb{R^+}$ and $S_{f,j}^t = \mathbb{K} \times \mathbb{R^+} \times \mathbb{R^+}$ respectively, where $\mathbb{K}=\mathbb{N} \cup \{0\}$.

Using the above-defined notation, the multi-period multi-leader-follower Stackelberg competition model can be expressed as follows:
\vspace{0mm}

\begin{align}
& \max_{s_l, s_f} \quad \Psi_l(s_l,s_f) =\sum_{i=1}^{N} \Psi_{l,i} (s_l,s_f) &\label{eq:lProfit}\\
& \hspace{1mm} \mbox{s.t.} & \nonumber\\
& \quad \quad s_f = {\argmax_{s_f}}\bm{\lbrace}
		\Psi_{f}(s_l,s_f) = \sum_{j=1}^{M} \Psi_{f,j} (s_l,s_f):\nonumber\\
& \hspace{57mm}		 q_{f,j}^{t} \geq Q_{f,j}^t , &\nonumber\\
&		 \hspace{57mm}  I_{f,j}^{t+1} \leq \alpha_f \! \left( P_{f,j}^t \,q_{f,j}^t - C_{f,j}^t \right) , &\nonumber\\
&		\hspace{57mm} M_{f,j}^{t+1} \leq \beta_f \! \left( P_{f,j}^t \,q_{f,j}^t - C_{f,j}^t \right) , &\nonumber\\ 
&		\hspace{57mm} I_{f,j}^1 = 0 , \quad M_{f,j}^1 = 0 .
\bm{\rbrace}&\\
& \quad \quad q_{l,i}^t \geq Q_{l,i}^t , &		\label{eq:lDemand}\\
& \quad \quad I_{l,i}^{t+1} \leq \alpha_l \! \left( P_{l,i}^t \,q_{l,i}^t - C_{l,i}^t \right) , &	\label{eq:lInvestment}\\
& \quad \quad M_{l,i}^{t+1} \leq \beta_l \! \left( P_{l,i}^t \,q_{l,i}^t - C_{l,i}^t \right) , &	\label{eq:lMarketing}\\
& \quad \quad I_{l,i}^1 = 0 , \quad M_{l,i}^1 = 0, & \label{eq:initial}\\
& \quad \quad s_{l} \in S_{l} , \quad s_{f} \in S_{f} & \label{eq:belongs}.
\end{align}

In the above formulation, it is assumed that all the leaders and followers are capable of satisfying the entire market's demand for their products, and this assumption is reflected by constraint \eqref{eq:lDemand} and the corresponding constraint for the followers. The restrictions on the size of investment and marketing expenditures are given by constraints (\ref{eq:lInvestment}, \ref{eq:lMarketing}) and the respective constraints for the followers. We also require that 
$0 \leq \alpha_{l}, \beta_l, \alpha_f, \beta_f \leq 1$, $(\alpha_l+\beta_l) \leq 1$ and $(\alpha_f+\beta_f) \leq 1$, which implies that investment and marketing expenditures must not exceed revenue in the previous period. Constraint \eqref{eq:initial} and the follower's respective constraint show that both investment and marketing are null for both firms in the first period because there is no revenue yet to fund these expenditures. 

The sum of profits of the leaders and followers may be further decomposed as
\[
\Psi_{l,i} (s_l,s_f) = \sum_{t=1}^T \Psi_{l,i}^t (s_{l},s_{f}) 
	\quad \mbox{and} \quad 
\Psi_{f,j} (s_l,s_f) = \sum_{t=1}^T \Psi_{f,j}^t (s_{l},s_{f}) ,
\]
where $\Psi_{l,i}^{t}$ and $\Psi_{f,j}^{t}$ are the profits generated by leader $i$ and follower $j$ during period $t$. The single period profits are calculated by deducting production, investment, and marketing expenses from total revenue, i.e.
\[
\Psi_{l,i}^t = P_{l,i}^t \,q_{l,i}^t - C_{l,i}^t - M_{l,i}^t - I_{l,i}^t 
	\quad \mbox{and} \quad
\Psi_{f,j}^t = P_{f,j}^t \,q_{f,j}^t - C_{f,j}^t - M_{f,j}^t - I_{f,j}^t .
\]

\section{Solution Methodology}\label{sec:method}
We solve the above problem using a computationally intensive nested scheme relying on the principles of evolutionary computation. The method is based on a steady state single-objective real-coded genetic algorithm to solve the problems at both levels. The underlying algorithm at both levels is a modified version of the procedures \cite{my-cec05,my-cec06} based on the single objective Parent Centric Crossover (PCX) \cite{pcx}. The nested bilevel evolutionary algorithm used to solve the described multi-period multi-leader-follower problem is a simple-minded strategy, where the lower level problem is solved for all given upper level decision vectors. Whenever the lower level optimal decision vector is to be determined for an upper level decision vector, the information is utilized from the nearest upper level decision vector for which the lower level optimal decision vector is known. A step-by-step procedure for the algorithm is described as follows:

\vskip 0.2cm
\textit{Step 1: Initialization.} A random population of size $N_p$ is initialized by generating the required number of upper level variables, and then the lower level optimization procedure is executed to determine the corresponding optimal lower level variables. Fitness is assigned based on upper level function value and constraints.

\textit{Step 2: Selection of upper level parents.} Randomly choose $2\mu$ members from the previously obtained population and conduct a tournament selection to determine $\mu$ sets of parents.

\textit{Step 3: Evolution at the upper level.} Create $\lambda$ offsprings by performing a PCX based crossover \citep{pcx} (Refer Sub-section \ref{sec:crossover}) and a polynomial mutation for the real variables, and a binary crossover and a binary mutation for the discrete variables.

\textit{Step 4: Lower level optimization.} For each of the generated offsprings, determine the closest upper level member in the population. Copy the lower level optimal variables from the closest upper level member. Thereafter, generate a lower level random population with a population size $n_p-1$. Include the lower level variables from the closest upper level member in the lower level population as the $n_{p}^{th}$ member. Perform the lower level optimization run with the generated population members.

\textit{Step 5: Evaluate offsprings.} Combine the upper level variables for the offsprings from Step 3 with the corresponding optimal lower level variables from Step 4, and evaluate each of the offsprings.

\textit{Step 6: Population update.} After evaluating the offsprings, choose $r$ random members from the parent population and pool them with the $\lambda$ offsprings. The best $r$ members from the pool replace the chosen $r$ members from the population.

\textit{Step 7: Termination check.} Perform a termination check and proceed to the next generation (Step 2) if the termination check (Refer Sub-section \ref{sec:termination}) is false.

\subsection{Lower Level Optimization}
The lower level optimization procedure is similar to the upper level procedure. The fitness assignment at this level is performed based on lower level function value and constraints. After initializing $n_p$ lower level members, $2\mu$ members are randomly chosen from the population. A tournament selection is performed and $\mu$ parents are chosen for crossover. Crossover and mutation operators are used to to generate $\lambda$ offsprings. A population update is performed as before by choosing $r$ random members from the population. A pool is formed using $r$ chosen members and $\lambda$ offsprings, from which the best $r$ members are used to replace the $r$ chosen members from the population. The next generation is executed if the termination criteria is not satisfied.

\subsection{Parameters}
The parameters in the algorithm are fixed as $\mu=3$, $\lambda=3$ and $r=2$. Crossover probability is fixed at $0.9$ and the mutation probability is $0.1$. The upper level population size $N_p$ and the lower level population size $n_p$ are fixed as 100.

\subsection{Crossover Operator}
\label{sec:crossover}
The crossover operator used in Step 3 is similar to the PCX operator proposed in \cite{my-cec06}. The operator creates a new solution from 3 parents as follows, choosing one of the parents as the index parent:
\begin{equation}
\mathbf{c} = \mathbf{x_p} + \omega_{\xi}\mathbf{d} + \omega_{\eta}\frac{\mathbf{p_2}-\mathbf{p_1}}{2}
\label{eq:child}
\end{equation}
The terms used in the above equation are defined as follows:
\begin{itemize}\itemsep1pt \parskip0pt \parsep0pt
	\item $\mathbf{x_p}$ is the {\em index\/} parent
	\item $\mathbf{d}=\mathbf{x_{p}}-\mathbf{g}$, where $\mathbf{g}$ is the mean of $\mu$ parents
	\item $\mathbf{p_1}$ and $\mathbf{p_2}$ are the other two parents
	\item $\omega_{\xi}=0.1$ and $\omega_{\eta}=\sum_{i=1}^{m_r} \frac{m_r}{|x_{p}^{i}-g^{i}|}$ are the two parameters, where $m_r$ is the number of real variables in the problem
\end{itemize}
The two parameters $\omega_{\xi}$ and $\omega_{\eta}$, describe the extent of variations along the respective directions. While creating $\lambda=3$ offsprings from $\mu=3$ parents, each parent is chosen as an index parent at a time.

\subsection{Constraint Handling}
The constraint handling scheme proposed by \citet{debpenalty} is used in this paper. At both levels, a constraint violation is defined as the sum of violations of all constraints for any point. A member of the population with a smaller constraint violation is always preferred by the algorithm over a member with a higher constraint violation. A member with no constraint violation is deemed to be a feasible, and is considered better than any of the other infeasible members. When making a comparison between two feasible members, the member with the higher function value is preferred.

\subsection{Termination Check}
\label{sec:termination}
The algorithm uses a variance-based termination criteria at both levels. When the value of $\eta$, described in the following equation becomes less than $\eta_{stop}$, the algorithm terminates.
\begin{equation}
\begin{array}{l}
	\eta = \sum_{i=1}^{m} \frac{\sigma^2(x_{current}^i)}{\sigma^2(x_{initial}^i)}.
\label{eq:criteria}
\end{array}
\end{equation}
The value of $\eta$ is restricted between 0 and 1 in (\ref{eq:criteria}). In the above equation, $m$ corresponds to the number of dimensions in the considered optimization problem. Thus, this value is equal to the number of lower level variables for the lower level problem and to the number of upper level variables for the upper level problem. The variables in the current population are denoted by $x_{current}^i : \mbox{i} \in \{1,2,\dotsc,m\}$, and the variables of the initial population are given by $x_{initial}^i : \mbox{i} \in \{1,2,\dotsc,m\}$. For both levels, the value of $\eta_{stop}$ is set as $10^{-5}$.

For all the runs performed in this paper, we execute a mutation-based local-search at both the levels after the termination of the algorithm to check if the algorithm has properly converged. In the later part of this paper, we evaluate the described nested bilevel evolutionary algorithm on a set of bilevel test problems, and then solve the multi-period multi-leader-follower problem.

\section{Algorithm Evaluation}\label{sec:evaluation}
The algorithm described above is evaluated on a set of recently proposed SMD \cite{my-cec12a} test problems to determine its efficacy towards handling difficult bilevel problems. The test set considered in this paper contains 6 test problems, which evaluate the solution procedure accross various difficulties offered by commonly encountered bilevel optimization problems. We provide the results of the algorithm on 10 and 20 variable instances of the SMD test problems. For 10-variable instances of the SMD problems, Table \ref{tab:SMDtable1} shows the number of function evaluations required at the upper and lower levels, and Table \ref{tab:SMDtable2} provides the accuracy\footnote{Accuracy is calculated by computing the Euclidean distance of the solution obtained using the algorithm from the actual optimal solution of the problem.} achieved along with the average number of lower level function evaluations required per lower level call. Similar results are reported for the 20-variable instances in Tables \ref{tab:SMDtable3} and \ref{tab:SMDtable4}. All performance metrics have been computed based on 31 runs of the evolutionary procedure.

\begin{table*}[!hbt]
\caption{Function evaluations (FE) for the upper level (UL) and the lower level (LL) from 31 runs of 10-variable instances of SMD test problems.} 
\label{tab:SMDtable1}
{\small\begin{center}
\begin{tabular}{|c|c|c|c|c|c|c|} \hline
Pr. No.	&	\multicolumn{2}{|c|}{Best}	&	\multicolumn{2}{|c|}{Median}	&	\multicolumn{2}{|c|}{Worst}	\\	\cline{2-7}
	&		\multicolumn{1}{|c|}{Total LL}	&	\multicolumn{1}{|c|}{Total UL}	&	\multicolumn{1}{|c|}{Total LL}	&	\multicolumn{1}{|c|}{Total UL}	&	\multicolumn{1}{|c|}{Total LL}	&\multicolumn{1}{|c|}{Total UL}	\\	
	&	\multicolumn{1}{|c|}{FE} 	&
        \multicolumn{1}{|c|}{FE}	&\multicolumn{1}{|c|}{FE}         &\multicolumn{1}{|c|}{FE}        &
        \multicolumn{1}{|c|}{FE}& \multicolumn{1}{|c|}{FE}	\\ \hline	
SMD1	&	853319	&	1150	&	1692741	&	2639	&	2178997	&	3640	\\	\hline
SMD2	&	1030789	&	1459	&	1557348	&	2387	&	2232534	&	3787	\\	\hline
SMD3	&	879785	&	1238	&	1476206	&	2390	&	1978647	&	2992	\\	\hline
SMD4	&	581078	&	710	&	1134376	&	1687	&	1438936	&	2153	\\	\hline
SMD5	&	1217102	&	1683	&	2021104	&	2966	&	2904080	&	3772	\\	\hline
SMD6	&	1303778	&	1646	&	2463058	&	3301	&	3230143	&	4165	\\	\hline
\end{tabular}
\end{center}}
\end{table*}

\begin{table*}[!hbt]
\caption{Accuracy for the upper and lower levels, and the lower level calls from 31 runs of 10-variable instances of SMD test problems.} 
\label{tab:SMDtable2}
\begin{center}
\begin{tabular}{|c|c|c|c|c|} \hline
Pr. No. & Median & Median & Median &  \\ \cline{2-4}
	& UL Accuracy & LL Accuracy & LL Calls &  $\frac{\mbox{LL Evals}}{\mbox{LL Calls}}$ \\ \hline
SMD1	&	0.000036	&	0.000015	&	2639	&	641.39	\\	\hline
SMD2	&	0.000006	&	0.000005	&	2387	&	652.39	\\	\hline
SMD3	&	0.000064	&	0.000023	&	2390	&	617.64	\\	\hline
SMD4	&	0.000028	&	0.000027	&	1687	&	672.60	\\	\hline
SMD5	&	0.000004	&	0.000003	&	2966	&	681.37	\\	\hline
SMD6	&	0.000157	&	0.000081	&	3301	&	746.07	\\	\hline
\end{tabular}
\end{center}
\end{table*}

\begin{table*}[!hbt]
\caption{Function evaluations (FE) for the upper level (UL) and the lower level (LL) from 31 runs of 20-variable instances of SMD test problems.} 
\label{tab:SMDtable3}
{\small\begin{center}
\begin{tabular}{|c|c|c|c|c|c|c|} \hline
Pr. No.	&	\multicolumn{2}{|c|}{Best}	&	\multicolumn{2}{|c|}{Median}	&	\multicolumn{2}{|c|}{Worst}	\\	\cline{2-7}
	&		\multicolumn{1}{|c|}{Total LL}	&	\multicolumn{1}{|c|}{Total UL}	&	\multicolumn{1}{|c|}{Total LL}	&	\multicolumn{1}{|c|}{Total UL}	&	\multicolumn{1}{|c|}{Total LL}	&\multicolumn{1}{|c|}{Total UL}	\\	
	&	\multicolumn{1}{|c|}{FE} 	&
        \multicolumn{1}{|c|}{FE}	&\multicolumn{1}{|c|}{FE}         &\multicolumn{1}{|c|}{FE}        &
        \multicolumn{1}{|c|}{FE}& \multicolumn{1}{|c|}{FE}	\\ \hline	
SMD1	&	2802304	&	2828	&	4665029	&	4686	&	6263539	&	6436	\\	\hline
SMD2	&	1839478	&	2568	&	3561479	&	3866	&	4694610	&	5304	\\	\hline
SMD3	&	3104455	&	2498	&	4988453	&	4494	&	6825706	&	6061	\\	\hline
SMD4	&	1287695	&	2069	&	2435796	&	3250	&	3362992	&	4434	\\	\hline
SMD5	&	4641337	&	4479	&	8933839	&	7404	&	12274821	&	9699	\\	\hline
SMD6	&	4959916	&	3481	&	8246468	&	6874	&	10985316	&	9043	\\	\hline
\end{tabular}
\end{center}}
\end{table*}

\begin{table*}[!hbt]
\caption{Accuracy for the upper and lower levels, and the lower level calls from 31 runs of 20-variable instances of SMD test problems.} 
\label{tab:SMDtable4}
\begin{center}
\begin{tabular}{|c|c|c|c|c|} \hline
Pr. No. & Median & Median & Median &  \\ \cline{2-4}
	& UL Accuracy & LL Accuracy & LL Calls &  $\frac{\mbox{LL Evals}}{\mbox{LL Calls}}$ \\ \hline
SMD1	&	0.000247	&	0.000143	&	4686	&	995.52	\\	\hline
SMD2	&	0.000146	&	0.000169	&	3866	&	921.23	\\	\hline
SMD3	&	0.000442	&	0.000155	&	4494	&	1110.03	\\	\hline
SMD4	&	0.000526	&	0.000171	&	3250	&	749.48	\\	\hline
SMD5	&	0.000011	&	0.000013	&	7404	&	1206.62	\\	\hline
SMD6	&	0.000745	&	0.000351	&	6874	&	1199.66	\\	\hline
\end{tabular}
\end{center}
\end{table*}

\begin{figure*}
\begin{minipage}[t]{0.48\linewidth}
\begin{center}
\epsfig{file=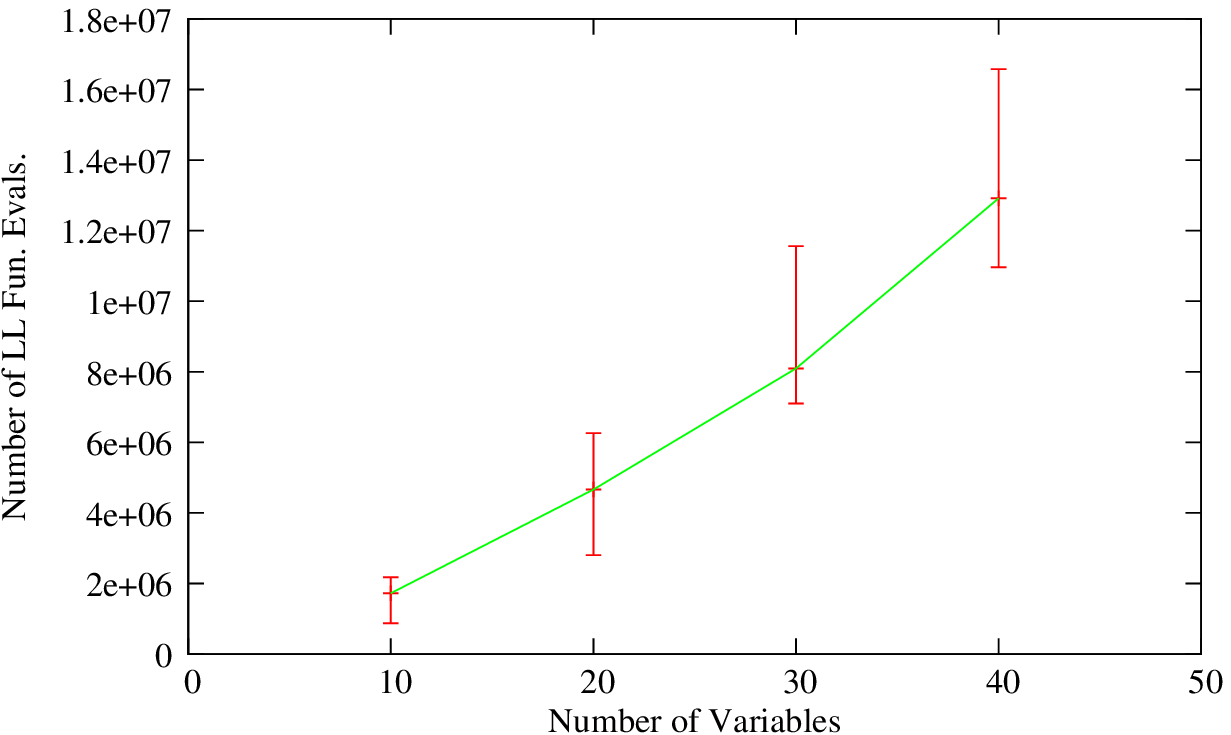,width=\linewidth}
\end{center}
\caption{Lower level function evaluations for SMD1 from 31 runs of the nested bilevel solution procedure.}
\label{fig:SMDfigure1}
\end{minipage}\hfill
\begin{minipage}[t]{0.48\linewidth}
\begin{center}
\epsfig{file=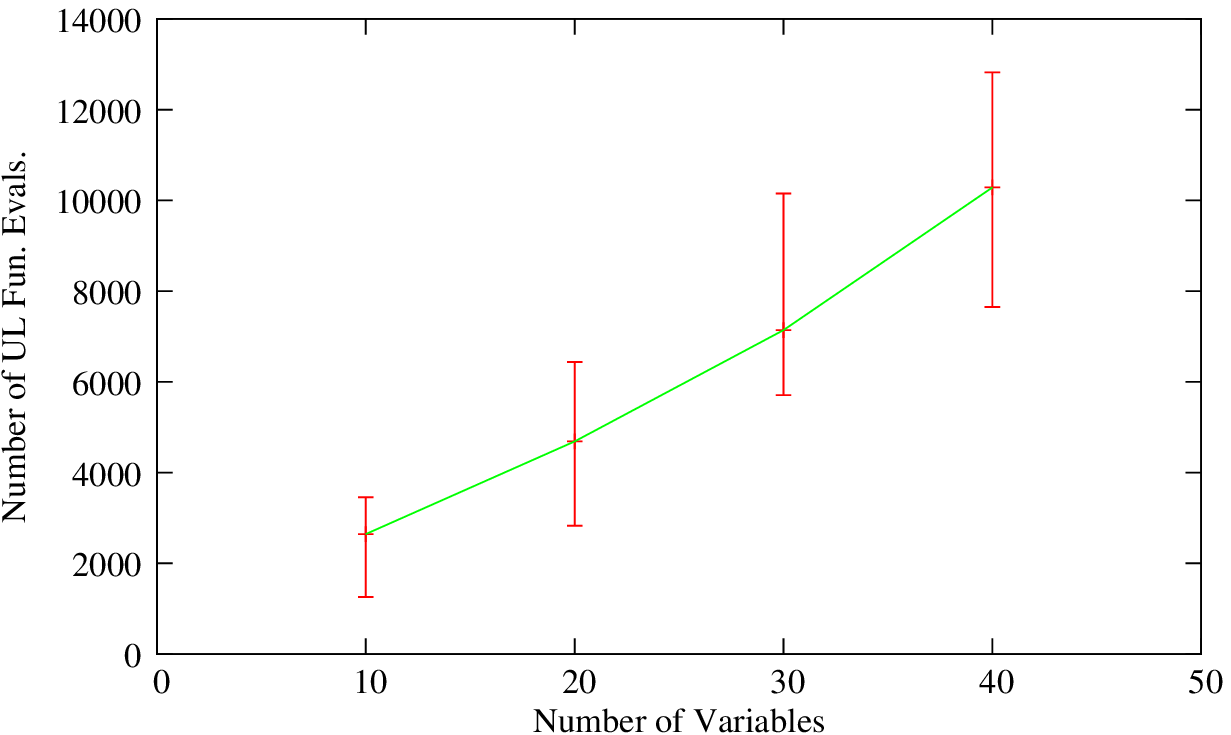,width=\linewidth} 
\end{center}
\caption{Upper level function evaluations for SMD1 from 31 runs of the nested bilevel solution procedure.}
\label{fig:SMDfigure2}
\end{minipage}
\end{figure*}

\begin{figure*}
\begin{minipage}[t]{0.48\linewidth}
\begin{center}
\epsfig{file=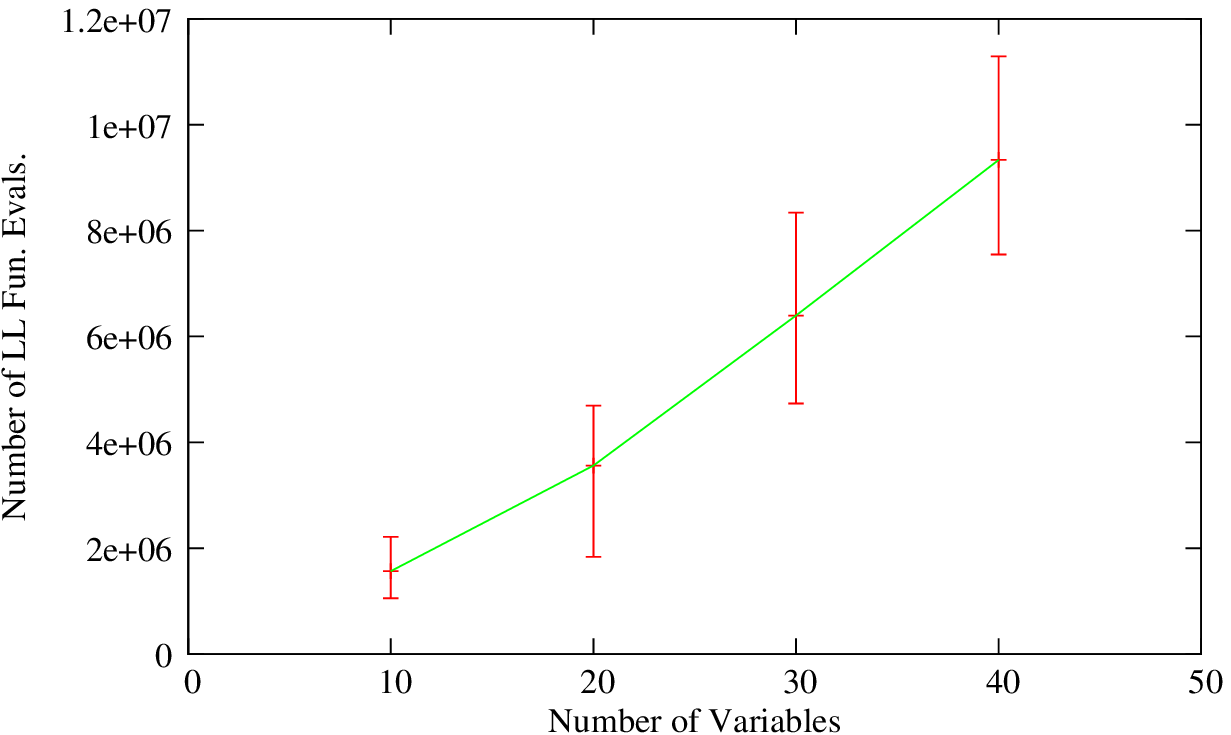,width=\linewidth}
\end{center}
\caption{Lower level function evaluations for SMD2 from 31 runs of the nested bilevel solution procedure.}
\label{fig:SMDfigure3}
\end{minipage}\hfill
\begin{minipage}[t]{0.48\linewidth}
\begin{center}
\epsfig{file=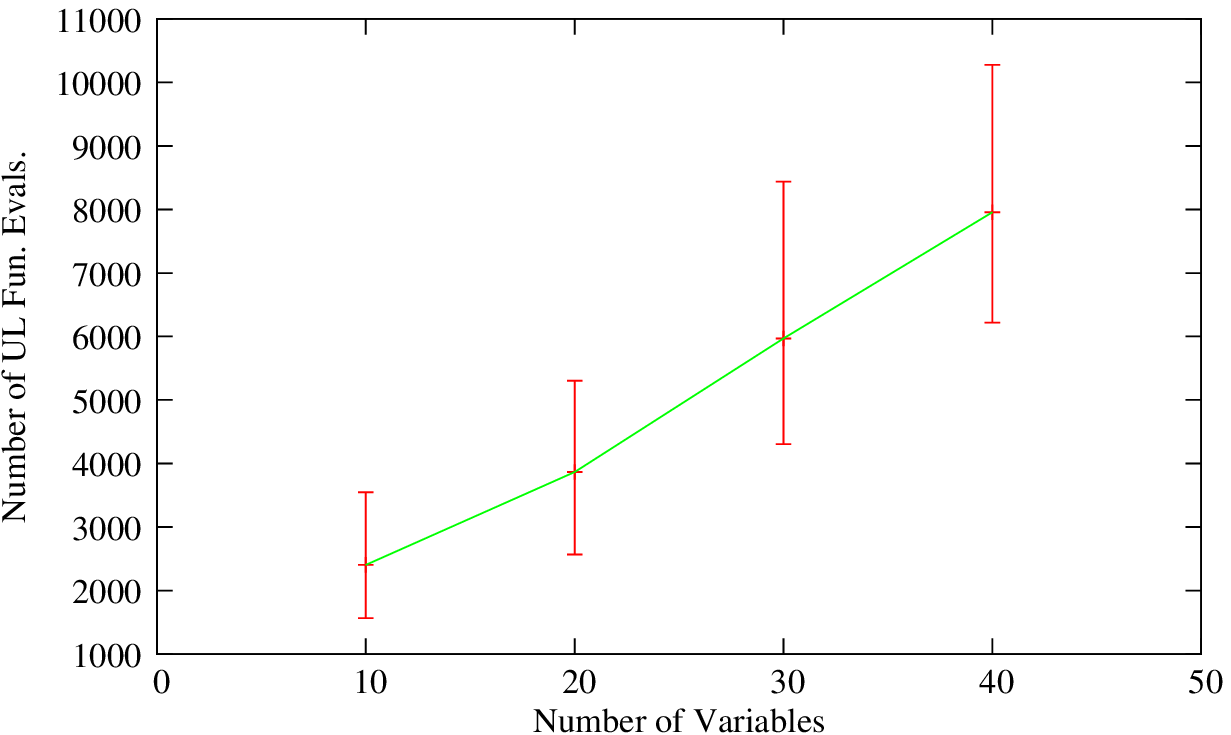,width=\linewidth} 
\end{center}
\caption{Upper level function evaluations for SMD2 from 31 runs of the nested bilevel solution procedure.}
\label{fig:SMDfigure4}
\end{minipage}
\end{figure*}

Next, we perform a scalability analysis to test the performance of the procedure, when the number of variables are even higher. SMD1 and SMD2 have been chosen for scalability study, and for both the problems 10, 20, 30 and 40 variable instances have been solved. For each instance, we have performed 31 runs, and the results for the required function evaluations are presented in Figures \ref{fig:SMDfigure1} and \ref{fig:SMDfigure2} for SMD1 and Figures \ref{fig:SMDfigure3} and \ref{fig:SMDfigure4} for SMD2. The algorithm converges successfully for each of the runs with a high accuracy.

\section{Results for Multi-Leader-Follower Stackelberg Model}
\label{sec:results}
In this section, we provide the cost and the inverse demand functions for the leaders and the followers in the multi-leader-follower Stackelberg model, and solve the problem using the nested bilevel evolutionary algorithm.
In the model, we assume that the $N$ leaders and $M$ followers are symmetric amongst themselves. Therefore, the cost and inverse demand functions for each firm in the respective groups are identical.
The cost and inverse demand functions for the leaders and followers at time period $t$ are defined as follows:
\begin{align}
	P_{l,i}^t &= \frac{100  \bigl(1+10{\sum_{r=1}^{t-1} M_{l,i}^{r}} \bigr)^{0.02}}
			{\bigl(1 + 0.01({\sum_{i=1}^{N} Q_{l,i}^{t}}+{\sum_{j=1}^{M} Q_{f,j}^{t}}) \bigr)^{0.5}} ,	\label{eq:lPrice}\\
	C_{l,i}^t &= \frac{0.025 q_{l,i}^2 + 0.02 q_{l,i} + 200 + 400 \log(1+q_{l,i})}
			{\displaystyle{\exp\!{\left(\frac{\sum_{r=1}^{t-1} I_{l,i}^{r}}{10,000}\right)}}} ,
\label{eq:lCost}\\
	P_{f,j}^t &= \frac{100  \bigl(1+10{\sum_{r=1}^{t-1} M_{f,i}^{r}} \bigr)^{0.02}}
			{\bigl(1 + 0.01({\sum_{i=1}^{N} Q_{l,i}^{t}}+{\sum_{j=1}^{M} Q_{f,j}^{t}}) \bigr)^{0.5}} ,	\label{eq:fPrice}\\
	C_{f,j}^t &= \frac{0.025 q_{f,i}^2 + 0.02 q_{f,i} + 200 + 400 \log(1+q_{f,i})}
			{\displaystyle{\exp\!{\left(\frac{\sum_{r=1}^{t-1} I_{f,i}^{r}}{10,000}\right)}}} . 	\label{eq:fCost}
\end{align}
The cost (\ref{eq:lCost},\ref{eq:fCost}) and inverse demand (\ref{eq:lPrice},\ref{eq:fPrice}) functions of the leaders and the followers are affected by their own cumulative investment and marketing expenditures respectively. Additionally, the price of the leaders' and followers' products depends on the total demand for the product and is inversely related to it. One could readily observe that we have kept the functions for the leaders similar to the followers even though no symmetry was assumed between the two groups. The rationale for this arrangement is discussed later in this section.

Cumulative investment introduces a further complication in the proposed model by creating an interaction between time periods. In other words, an amount invested in period $t$ will affect not only period $t+1$ but also all of the following periods. This prevents us from solving the model one period at a time. Thus, when considering each period, the investment in all previous periods must be taken into account. Cumulative marketing also leads to interaction among time periods. Even though the two group of firms have similar price functions, their products need not be homogeneous, as the price not only depends on the overall demand but is also influenced by the marketing expenditures made by the firms. 

Investment may also be viewed as a production adjustment cost, if the company is investing in additional capacity or improving its capital stock. \citet{Jun04} studied a symmetric duopoly with production adjustment costs. These costs may be modeled in different ways. For example, production adjustment costs may include menu costs, setup costs, and per-unit cost changes due to economies of scale. While we do not view investment as an adjustment cost in our presented model, it may still be considered as one due to our assumption of the firms investing in R\&D or new capital.

As explained by \citet{Karray09}, the marketing effects of competing firms that produce substitutable products are ambiguous. The effectiveness of marketing depends largely on the availability of media outlets. When information is readily available, marketing tends to be complimentary in the sense that one firm's marketing increases the demand not only for its own products, but also for similar products made by other firms. Conversely, with few media outlets, marketing is competitive and mostly increases the demand for the marketing firm's products. Based on this, the demand curves of each firm should ideally include the effects of its own marketing as well as that of the rival firm. However, since the effects of rival advertising can be either positive or negative, further assumptions are required to properly include them into the model. We do not make any such assumptions and exclude the rival firm's marketing effect from the inverse-demand function, thereby assuming that any possible complimentary effects of rival marketing are offset by the competitive aspects.

Next, we present the results obtained using the nested bilevel evolutionary algorithm on the multi-period multi-leader-follower problem. We solve the model described in equations \eqref{eq:lProfit}-\eqref{eq:belongs}, with the profit and cost functions defined in \eqref{eq:lPrice}-\eqref{eq:fCost}. Since we assume symmetrical leaders and symmetrical followers, the decision variables are identical among the leaders as well as among the followers at the optimum. It should be noted that the nature of the cost and inverse demand functions assumed for the leaders is identical to that of the followers. However, as we will observe from the results, the decision variables for the leaders will be different from the followers at the optimum. The profit and cost functions of the leaders have been kept similar to the followers for the purpose of highlighting a very important and defining characteristic of Stackelberg competition: the first mover's advantage experienced by the leaders over the followers. The results in this section are divided into two parts. In the first part, we present the results for a particular scenario, where there are 2 leaders and 5 followers. Thereafter, in the second part, we analyse the multi-leader-follower problem by varying the leaders and followers in the model.

\subsection{Part 1}
In this sub-section, we discuss the results obtained using the nested bilevel evolutionary algorithm on the multi-leader-follower problem with 2 leaders and 5 followers. The considered problem has been optimized over $5$ time periods. There are 20 variables for an individual firm in the Stackelberg framework, with 5 corresponding to demand, 5 corresponding to production, 5 corresponding to investment, and 5 corresponding to marketing. This makes the overall problem a $140$-variable problem with $40$ variables at the upper level and $100$ variables at the lower level. Out of these variables, the variables corresponding to production and demand are discrete, and the rest are continuous. The variable bounds for the discrete variables are all integer values from 0 to 1000, and the variable bounds for the continuous variables are all real values from 0 to 1000. 

We do not solve the problem as is, rather we first simplify it based on ideas from microeconomic theory:
\begin{itemize}\itemsep1pt \parskip0pt \parsep0pt
\item At the optimum, demand is equal to supply,
\item When firms are symmetric, their behaviour at Cournot equilibrium is identical.
\end{itemize} 
Demand being equal to supply at the optimum eliminates the demand variables from the model, which reduces the model size by 5 variables for an individual firm. Furthermore, the first period always has zero investment and marketing expenditures at both levels, which allows us to eliminate 2 more variables for each firm. Therefore, we finally end up with 13 variables for every firm in the model. Our assumption of symmetrical leaders and symmetrical followers reduces the size of the optimization problem being solved to 26, where the upper level optimization problem has $13$ variables and the lower level problem also has $13$ variables. Both levels contain an investment and a marketing constraint for each of the time periods (excluding the first time period) for each of the firms. This makes the number of constraints at the upper level as $8$ and the number of constraints at the lower level as $8$.


From here onwards, we present the results obtained by solving the bilevel optimization model using the nested bilevel evolutionary algorithm. Multiple runs were performed, each starting with a different random initial population. We chose the best solution obtained from 31 runs to report the results for the leader-follower problem. The upper and lower level function evaluations required by the algorithm to converge towards the optimal solution are shown in Table~\ref{tab:table0}.
One could readily observe the high number of lower level function evaluations required to solve the multi-leader-follower problem. The reason for this, as mentioned earlier, is that the nested approach solves a lower level optimization task for each new upper level member generated by the algorithm. 
\begin{table}
\caption{Computational expense required by the nested bilevel evolutionary algorithm to solve the multi-period multi-leader-follower problem with 2 leaders and 5 followers.}
\begin{center}
    \begin{tabular}{|c|c|c|c|}
        \hline
Function Evaluations   &   Upper Level         &  Lower Level\\ \hline
Min &   3,564  &  39,356,642 \\
Median &   4,610  & 55,553,874 \\
Max & 6,324 & 76,456,562 \\
\hline
    \end{tabular}
\end{center}
\label{tab:table0}
\end{table}

\begin{figure*}
\begin{minipage}[t]{0.49\linewidth}
\begin{center}
\epsfig{file=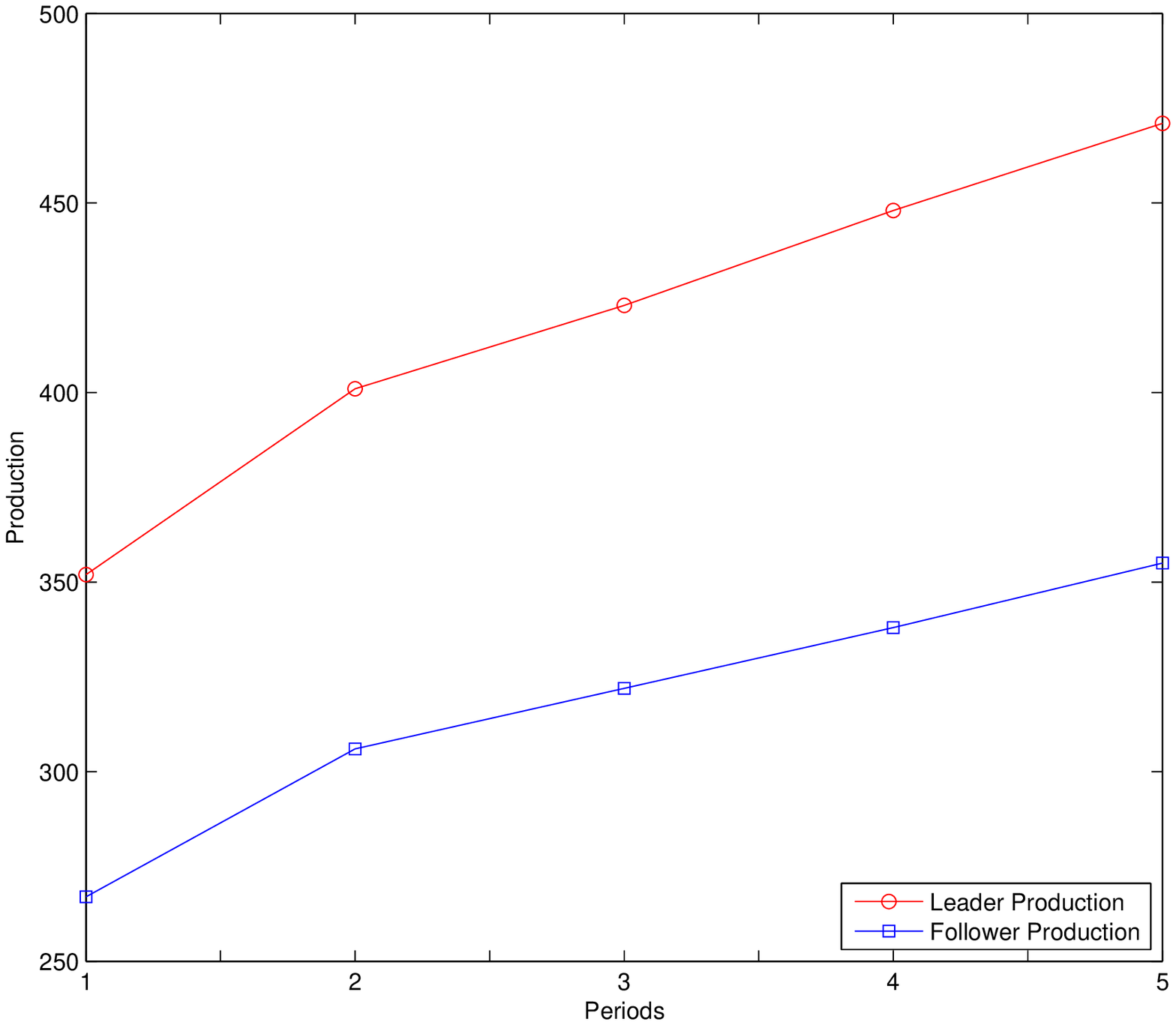,width=\linewidth}
\end{center}
\caption{Leader and follower production in a multi-leader-follower model with 5 time periods.}
\label{fig:figure1}
\end{minipage}\hfill
\begin{minipage}[t]{0.49\linewidth}
\begin{center}
\epsfig{file=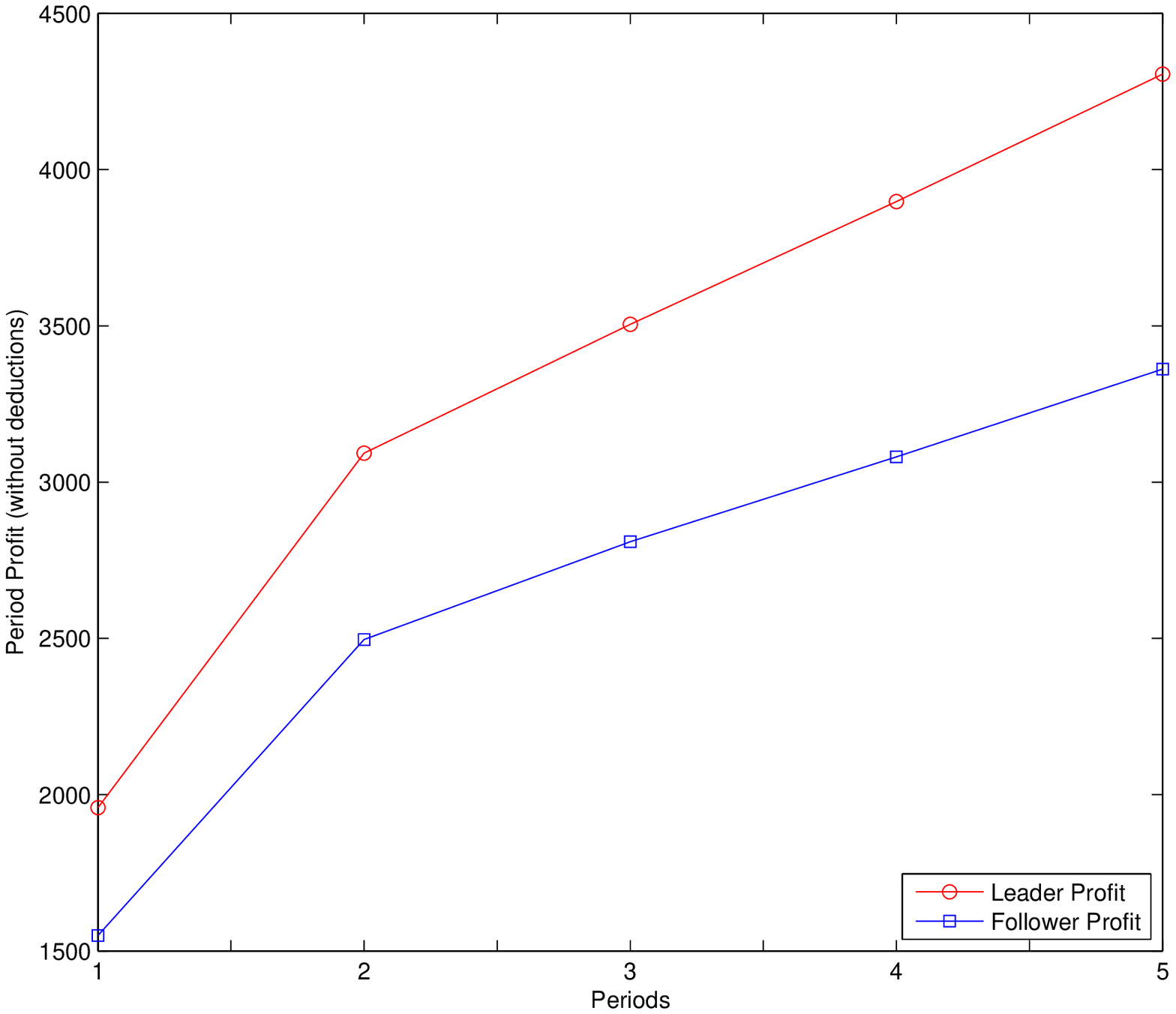,width=\linewidth} 
\end{center}
\caption{Leader and follower profits in each of the time-periods. Investment and marketing expenditures are not deducted from the reported profits.}
\label{fig:figure2}
\end{minipage}
\end{figure*}

\begin{figure*}
\begin{minipage}[t]{0.49\linewidth}
\begin{center}
\epsfig{file=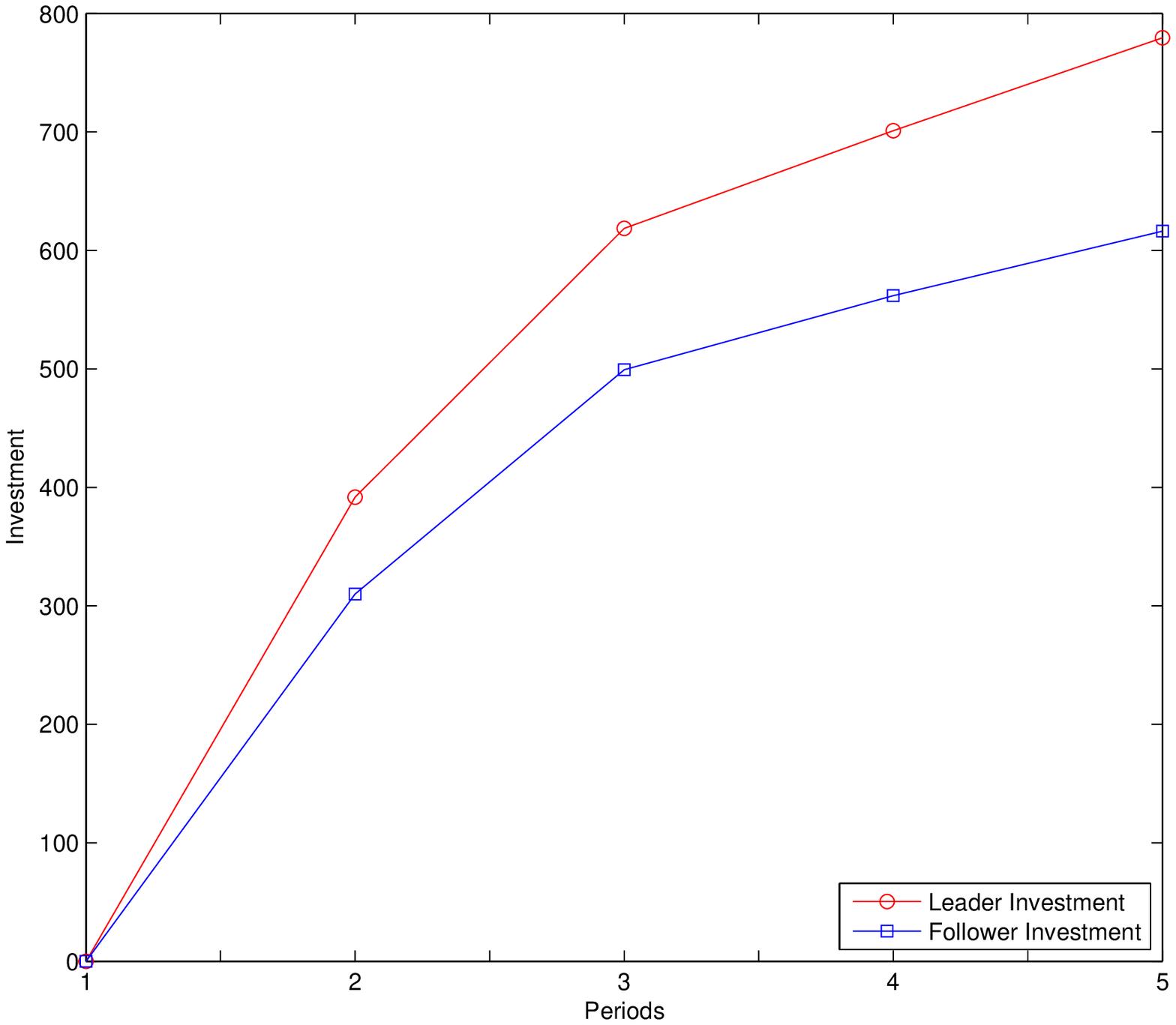,width=\linewidth} 
\end{center}
\caption{Leader and follower investment expenditures in a Stackelberg competition model with 5 time periods.}
\label{fig:figure3}
\end{minipage}\hfill
\begin{minipage}[t]{0.49\linewidth}
\begin{center}
\epsfig{file=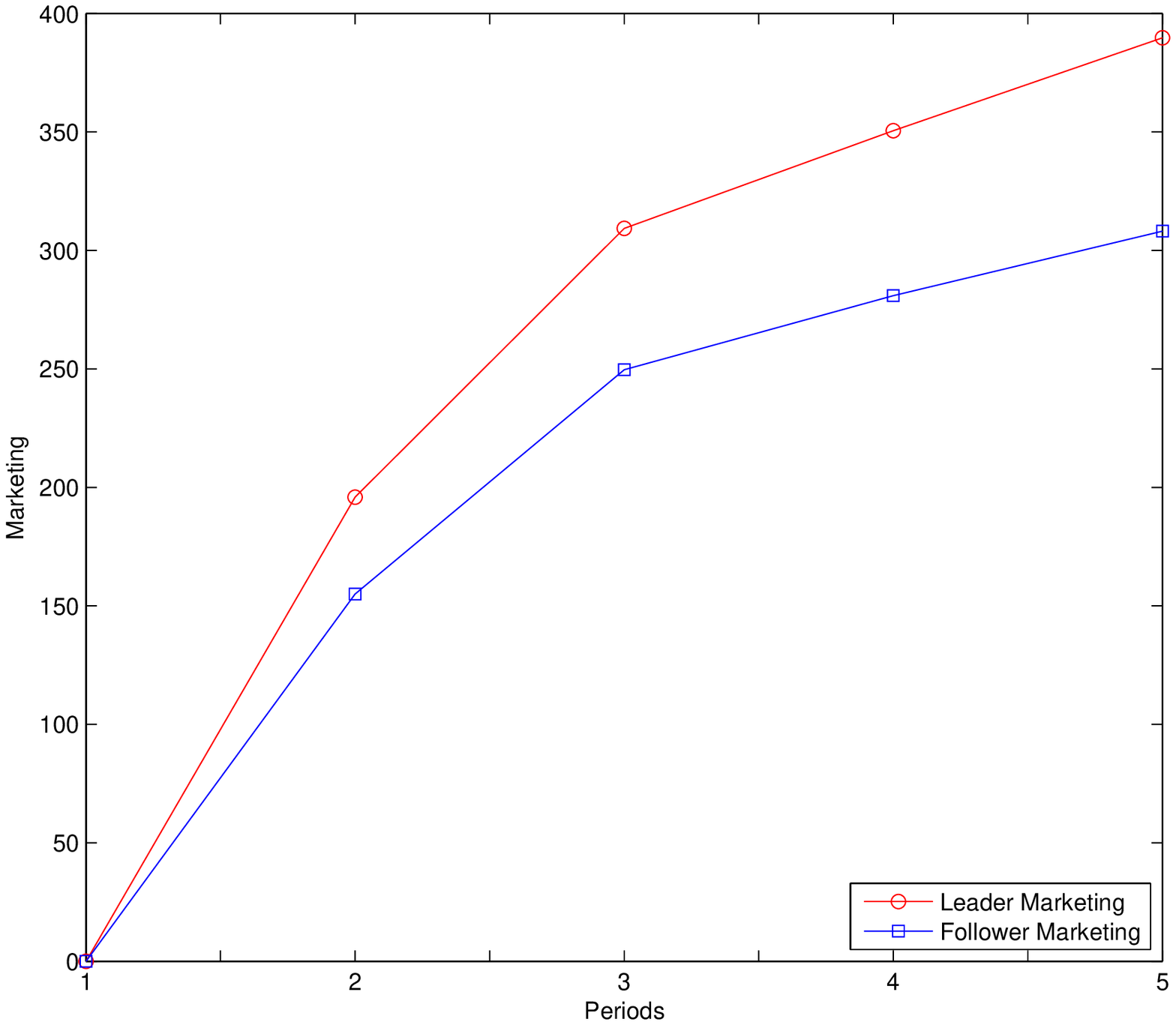,width=\linewidth}
\end{center}
\caption{Leader and follower marketing expenditures in a Stackelberg competition model with 5 time periods.}
\label{fig:figure4}
\end{minipage}
\end{figure*}

The best results from 31 runs are shown in Figures~\ref{fig:figure1},~\ref{fig:figure2},~\ref{fig:figure3} and~\ref{fig:figure4}. Figures~\ref{fig:figure1} and~\ref{fig:figure2} present the optimal production and period profits (without deductions) over multiple time periods for each of the firms in the system. The period profits are reported without deducting the marketing and investment expenditures made by the firm.
The graphs show that the leaders produce more and make a higher profit as compared to the followers. The difference is due to the first mover's advantage of the leaders. Figures~\ref{fig:figure3} and~\ref{fig:figure4} show the investment and marketing expenditures for the leaders and the followers. All of the constraints in the problem under consideration are active at the optima, thus keeping the investment and marketing variables at their upper limits in the constraints. A further relaxation of the constraints would lead to an increase in profits.

\begin{figure*}
\begin{center}
\epsfig{file=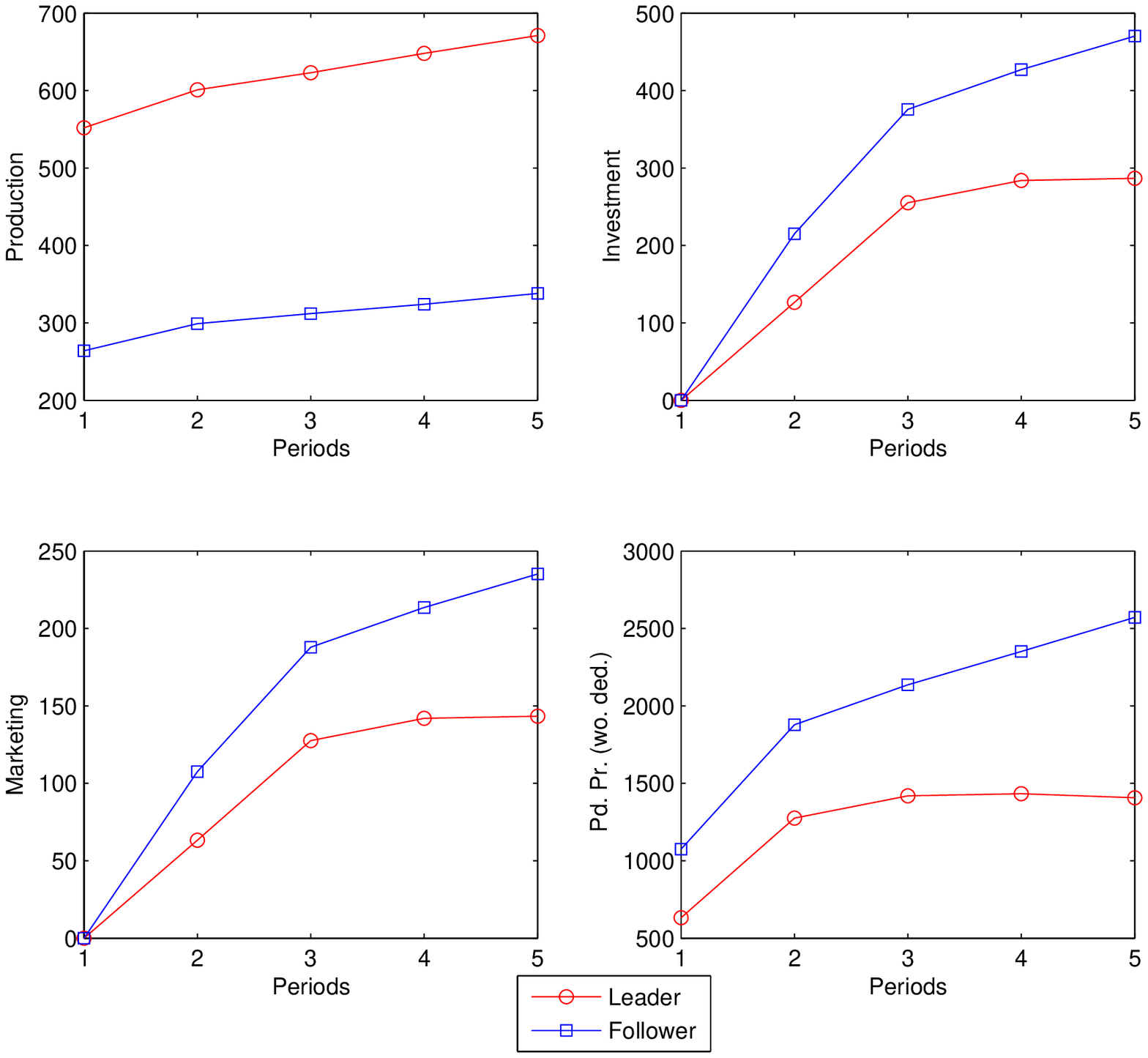,width=.67\linewidth}
\caption{Leader's objective function with respect to non-optimal production variables, when the follower follows an optimal strategy for each upper level vector.}
\label{fig:revisionNonOptimal1}
\end{center}
\end{figure*}
\begin{figure*}
\begin{center}
\epsfig{file=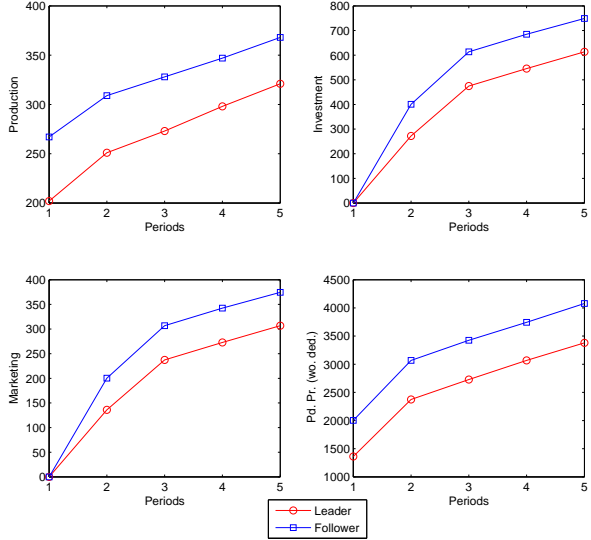,width=.67\linewidth}
\caption{Leader's objective function with respect to non-optimal production variables, when the follower follows an optimal strategy for each upper level vector.}
\label{fig:revisionNonOptimal2}
\end{center}
\end{figure*}

After presenting the optimal solution to the problem, we look at a couple of non-optimal solutions. We consider two non-optimal upper level decisions, and observe the optimal lower level responses to those decisions. For simplicity, it is assumed that all the leaders operate with the same non-optimal strategy. Firstly, we consider a scenario, where the leaders decide to produce substantially more than what is suggested by the Stackelberg optimum, while utilizing the maximum possible budget for investment and marketing. Figure \ref{fig:revisionNonOptimal1} shows this scenario, and the resulting profits of the leaders and the followers. It can be observed that under such a situation, the leaders make substantially lower profits, and the profits of the followers also get reduced. Individual leaders end up making smaller profits than the followers at this non-optimal point even though their productions are higher. Secondly, we consider a scenario, where the leaders decide to produce substantially less than what is suggested by the Stackelberg optimum, while utilizing the maximum possible budget for investment and marketing. Figure \ref{fig:revisionNonOptimal2} shows this scenario, and the resulting profits of the leaders and the followers. In this scenario, the leaders once again make substantially lower profits, and the profits of the followers rise significantly. 

\begin{figure*}
\begin{center}
\epsfig{file=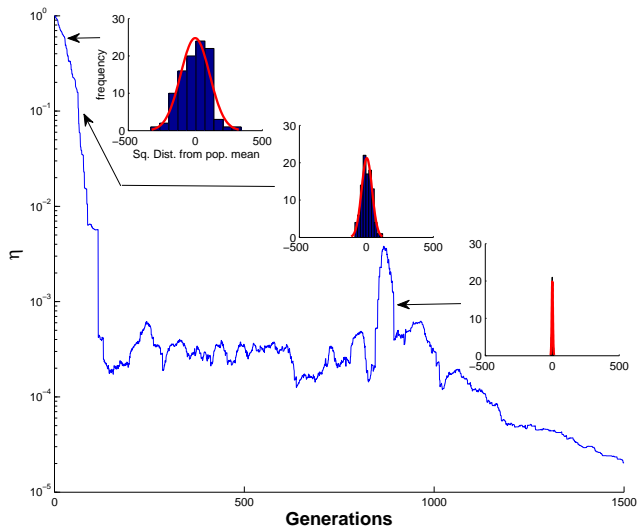,width=0.65\linewidth}
\caption{Termination check, $\eta$, over generations. The algorithm terminates when $\eta\le\eta_{stop}$. Population distributions are shown at three different stages of the algorithm.}
\label{fig:alphaConvergence}
\end{center}
\end{figure*}

Next, we look at the termination check, $\eta$, over generations from a particular run of the algorithm. It can be observed in Figure \ref{fig:alphaConvergence} that the value of $\eta$ drops sharply over the initial generations of the algorithm and then the algorithm slowly converges. The distribution of the population members at various stages of the algorithm are also shown in the figure.

The lower level optimization problem is a mixed integer non-linear programming problem. Therefore, it is possible to handle the lower level problem using a MINLP Solver. We replace the lower level evolutionary scheme implemented in the nested bilevel evolutionary algorithm with a MINLP Solver (TOMLAB Optimization) and record the savings in function evaluation. Table \ref{tab:tableMINLP} provides the results, when a MINLP Solver is used at the lower level. One can observe that using this approach leads to savings of almost $60\%$ at the lower level. However, it is noteworthy that the MINLP Solver makes the procedure less generic, as it cannot be used at the lower level in case of lower level problems having complications like multi-modalities, and various kinds of discontinuities. For example, the solver will fail to handle even small instances of SDM3, SMD4 and SMD5 test problems.
\begin{table}
\caption{Computational expense required by the MINLP Solver to solve the multi-period multi-leader-follower problems with 2 leaders and 5 followers.}
\begin{center}
    \begin{tabular}{|c|c|c|c|}
        \hline
Function Evaluations   &   Upper Level         &  Lower Level\\ \hline
Min &   3,468  &  16,458,426 \\
Median &   4,584  & 22,663,442 \\
Max & 6,782 & 36,763,428 \\
\hline
    \end{tabular}
\end{center}
\label{tab:tableMINLP}
\end{table}

\subsection{Part 2}
In this sub-section, we further analyze the problem by varying the number of leaders and followers in the system. We vary the number of leaders as well as followers from 1 to 5, which generates a total of 25 different Stackelberg competition scenarios. Each competition model is separately optimized and the results are reported. The cost and the inverse-demand functions are all kept fixed throughout the simulation study. The model is the same as before except that the number of leaders and followers are varying.

\begin{figure*}
\begin{minipage}[t]{0.49\linewidth}
\begin{center}
\epsfig{file=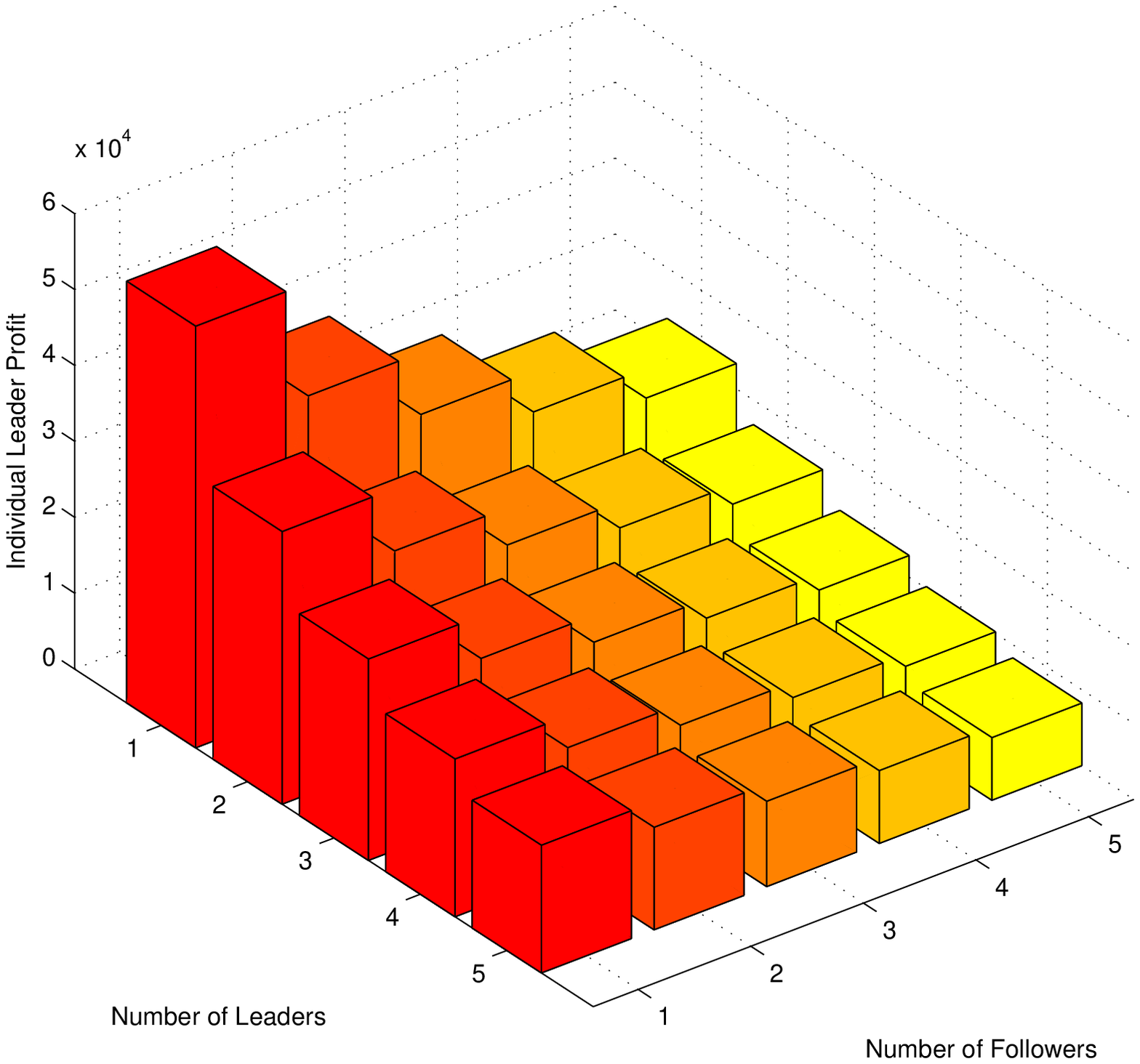,width=\linewidth} 
\end{center}
\caption{Individual leader profit (cumulative profit in 5 time periods) in a multi-leader-follower model when the number of leaders and followers are varied.}
\label{fig:figure5}
\end{minipage}\hfill
\begin{minipage}[t]{0.49\linewidth}
\begin{center}
\epsfig{file=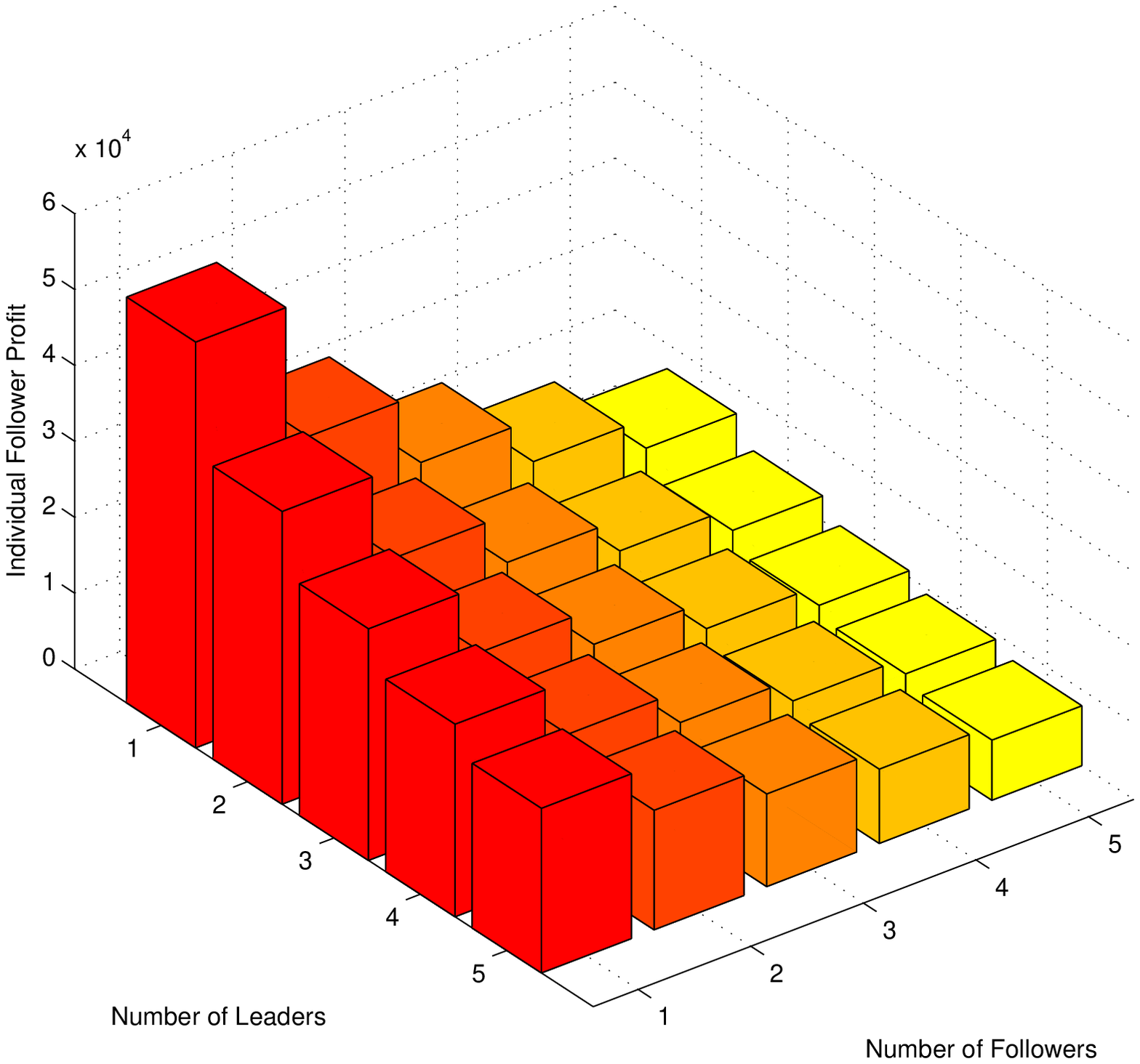,width=\linewidth}
\end{center}
\caption{Individual follower profit (cumulative profit in 5 time periods) in a multi-leader-follower model when the number of leaders and followers are varied.}
\label{fig:figure6}
\end{minipage}
\end{figure*}

\begin{figure*}
\begin{minipage}[t]{0.49\linewidth}
\begin{center}
\epsfig{file=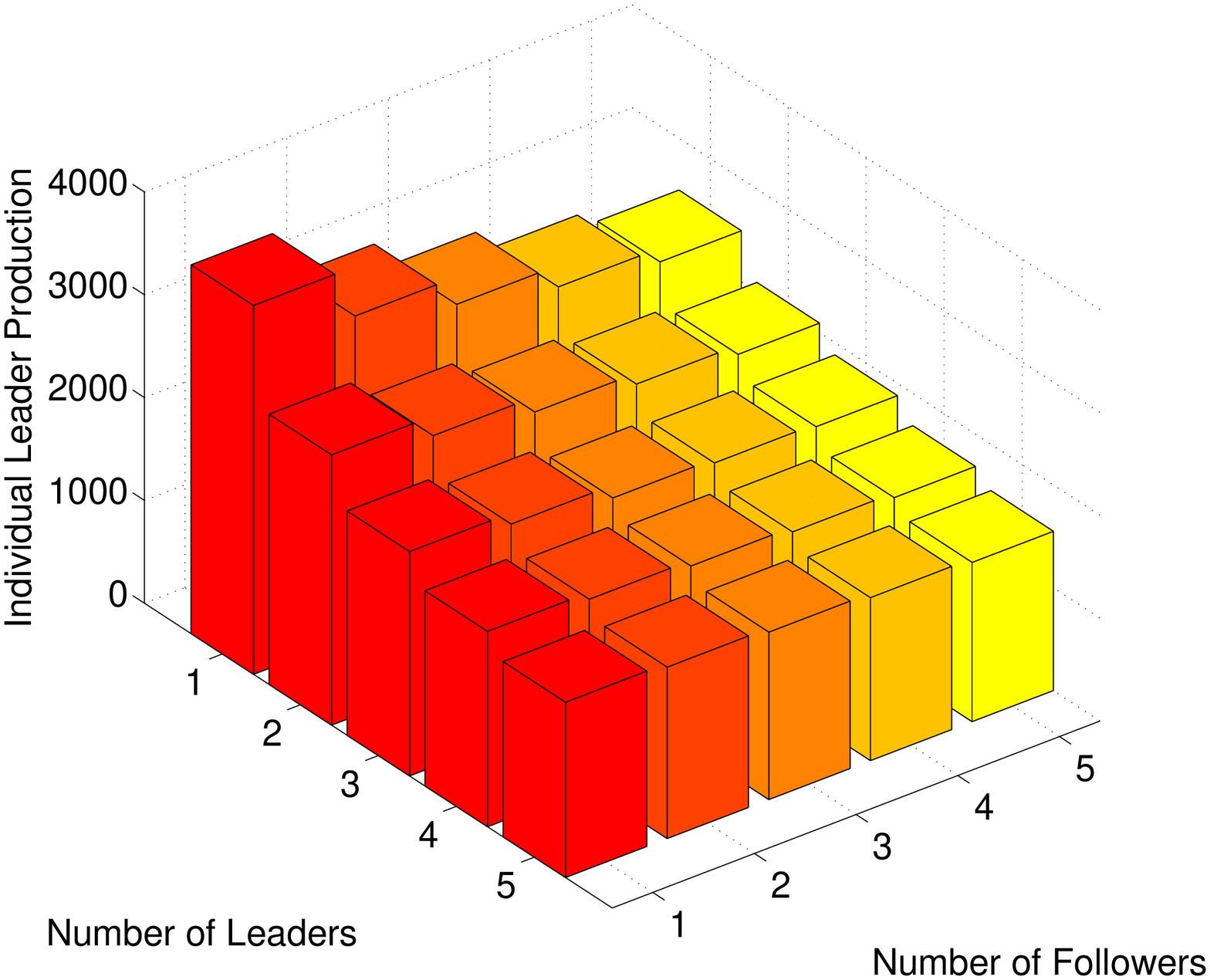,width=\linewidth}
\end{center}
\caption{Individual leader production levels (cumulative productions in 5 time periods) in a multi-leader-follower model when the number of leaders and followers are varied.}
\label{fig:figure7}
\end{minipage}\hfill
\begin{minipage}[t]{0.49\linewidth}
\begin{center}
\epsfig{file=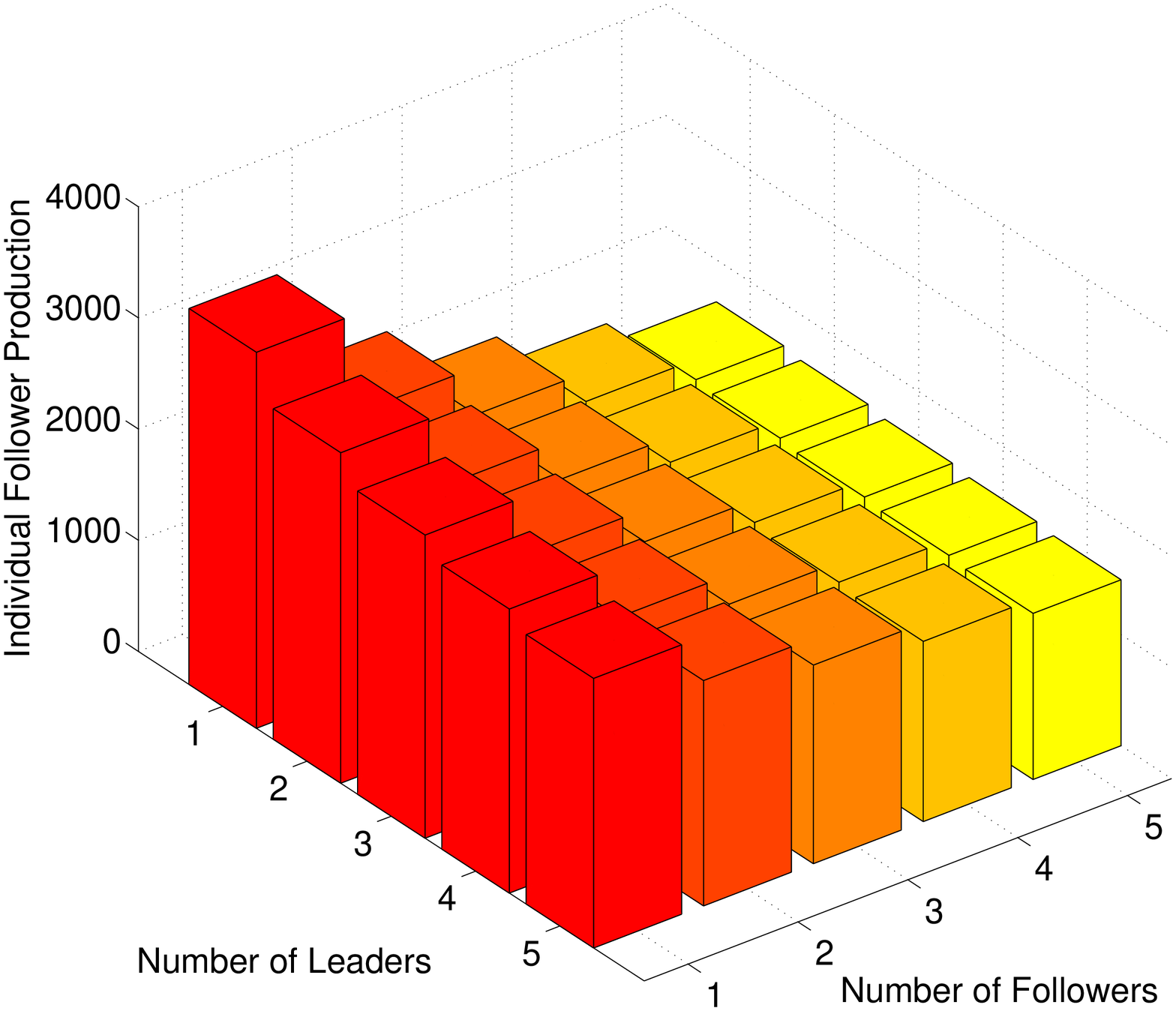,width=\linewidth} 
\end{center}
\caption{Individual follower production levels (cumulative productions in 5 time periods) in a multi-leader-follower model when the number of leaders and followers are varied.}
\label{fig:figure8}
\end{minipage}
\end{figure*}

Figures~\ref{fig:figure5} and~\ref{fig:figure6} show the profit (without deductions) for individual leaders and followers. We observe that in the case of a single leader and a single follower, the leader makes a higher profit as compared to the follower. The higher profit for the leader, in this case, can be attributed to the first mover's advantage. In the cases, when the number of leaders or followers is increased, the profit at the corresponding level gets distributed among the players. Since the leaders are identical, they make the same profit, and this is also the case for the followers. We observe that an addition of a player either at the leaders' level or at the followers' level causes a reduction in profit for all of the players. The reduction is substantial for the existing leaders, if a new player enters at the leaders' level, vis-{\`{a}}-vis, if a new player enters at the followers' level, it causes a substantial reduction in profits for the existing followers. Figures~\ref{fig:figure7} and~\ref{fig:figure8} show the production levels for the leaders and followers respectively, and we observe that the production levels for the individual players are falling with every new entrant.

Table~\ref{tab:table1} provides the values for total profit, without deduction of marketing and investment expenditures, generated from the market in each of the models. It is interesting to observe that the total profit, i.e. the sum of profits earned by all of the players in the model, reduces with an increasing number of players. On the other hand, if we look at the production levels in Table~\ref{tab:table2}, we observe that the total production increases with increasing number of players in the model. From this, we conclude that each extra player, either at the upper level or at the lower level, leads to an increase in production and a reduction in price to such an extent that the overall profit for the players falls. This offers an advantage to the consumers as the market becomes more competitive.
\begin{table}[t]
\caption{Sum of individual profits without deductions over all the players at the optimum for different Stackelberg competitions with $i$ number of leaders and $j$ number of followers. The number of leaders is presented across rows and the number of followers across columns.}
\begin{center}
    \begin{tabular}{|c|c|c|c|c|c|}
        \hline
$10^5 \times$   &   1 F         &  2 F         &  3 F         &  4 F         &  5 F      \\ \hline
1 L &   1.0900&    1.1133&    1.1122&    1.0984&    1.0695\\
2 L &  1.1038&    1.0892&    1.0690&    1.0386&    1.0008\\
3 L &  1.1013&    1.0634&    1.0339&    0.9847&    0.9294\\
4 L &  1.0849&    1.0306&    0.9800&    0.9270&    0.8719\\
5 L &  1.0573&    0.9898&    0.9287&    0.8700&    0.8088\\
        \hline
    \end{tabular}
\end{center}
\label{tab:table1}
\vspace{0mm}
\end{table}

\begin{table}[t]
\caption{Sum of individual productions over all the players at the optimum for different Stackelberg competitions with $i$ number of leaders and $j$ number of followers. The number of leaders is presented across rows and the number of followers across columns.}
\begin{center}
    \begin{tabular}{|c|c|c|c|c|c|}
        \hline
$10^4 \times$   &   1 F          &   2 F          &   3 F          &   4 F          &   5 F          \\ \hline
1 L&   0.6970 &   0.8087 &   0.9033 &   0.9843 &   1.0597 \\
2 L&   0.8231 &   0.9513 &   1.0500 &   1.1362 &   1.2122 \\
3 L&   0.9271 &   1.0612 &   1.1563 &   1.2533 &   1.3404 \\
4 L&   1.0158 &   1.1543 &   1.2631 &   1.3557 &   1.4348 \\
5 L&   1.0945 &   1.2388 &   1.3504 &   1.4406 &   1.5210 \\
        \hline
    \end{tabular}
\end{center}
\label{tab:table2}
\end{table}

\section{Convergence Analysis}\label{sec:convergence}
In this section, we analyse the convergence properties of the nested bilevel evolutionary algorithm on a single-leader and single-follower model operating in two time periods. The reason for choosing a simple problem with two time periods is that it would involve two production variables for the leader as well as the follower making it possible for us to represent the solutions graphically. Moreover, it also enables the use of a grid search around the optimal solution found by the evolutionary algorithm to check if the algorithm has converged to the true optima.

\begin{figure*}
\begin{center}
\epsfig{file=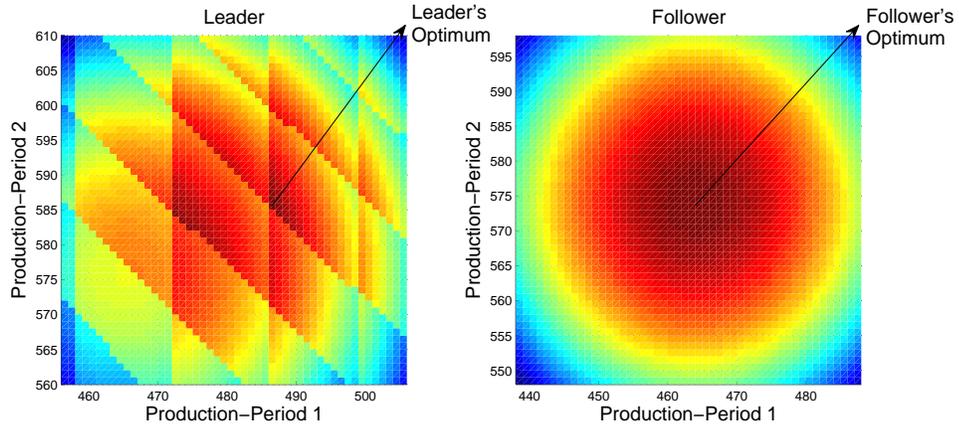,width=\linewidth}
\caption{{\em Left:} Leader's profit contour for different upper level decision vectors when the follower follows an optimal strategy for each upper level vector. {\em Right:}  Follower's profit contour for different lower level decision vectors at the optimal upper level decision vector.}
\label{fig:upperLower}
\end{center}
\end{figure*}

To begin with, we execute the nested bilevel evolutionary algorithm on this simple problem. The solution obtained by the approach is given in Table \ref{tab:simple}. A grid search is then performed around the obtained solution at the upper and the lower levels to verify the veracity of the solution obtained using the nested bilevel evolutionary algorithm. The solution of the grid search is shown in Figure \ref{fig:upperLower} with respect to the production variables. A lower level grid search was performed for each of the upper level vectors, and the lower level optimal solutions corresponding to those upper level vectors were determined. The left panel shows the upper level profit contours, when the lower level decision vectors are at their optimum. The right panel shows the lower level profit contours for one particular upper level decision vector, which is the optimal upper level decision vector. The optimal solution obtained using the grid search is found to be the same as the one obtained using the nested bilevel evolutionary approach. All the constraints were found to be active at the optimum.

\begin{figure*}
\begin{minipage}[t]{0.49\linewidth}
\begin{center}
\epsfig{file=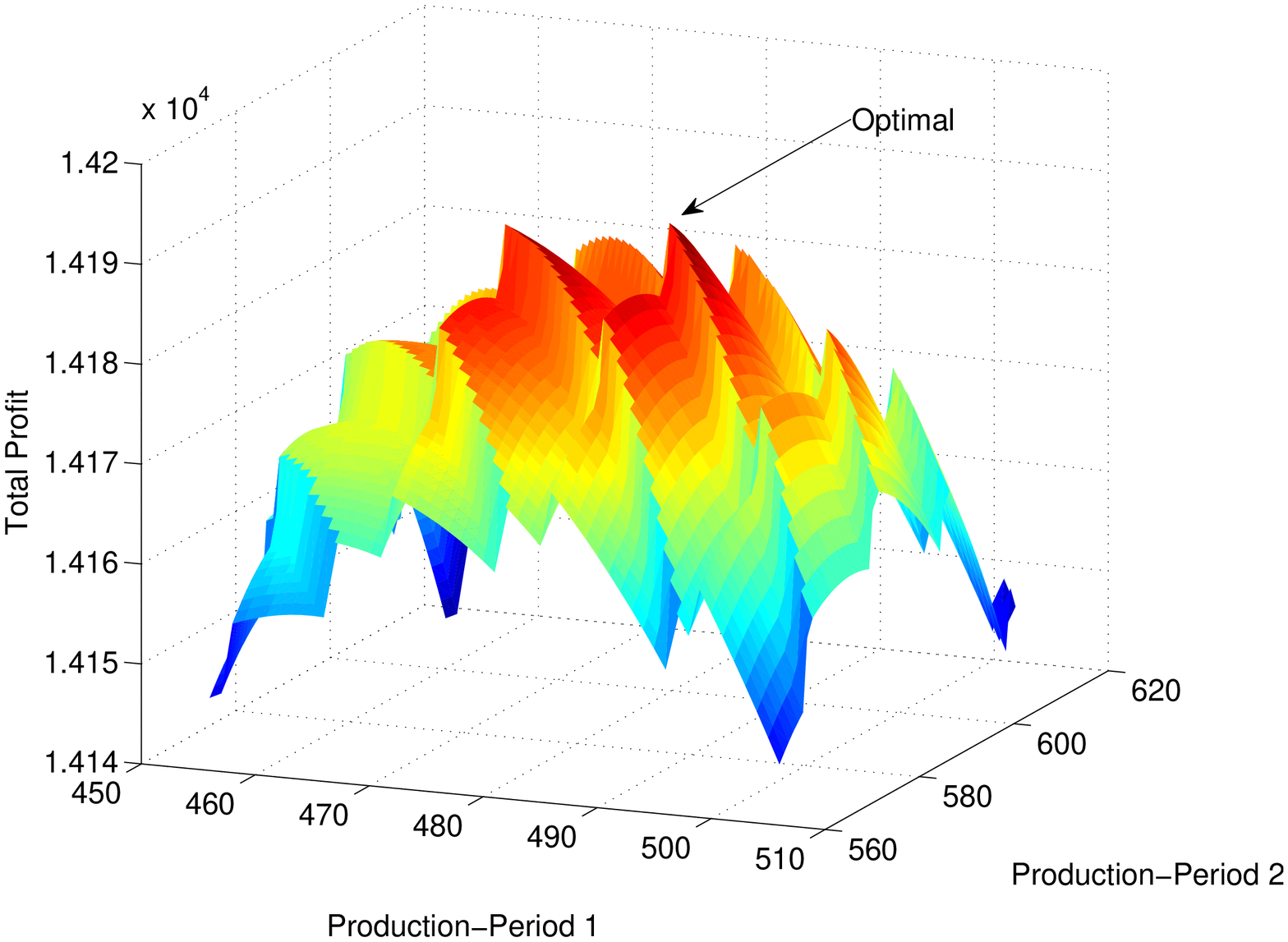,width=\linewidth}
\end{center}
\caption{Leader's objective function with respect to production variables, when the follower follows an optimal strategy for each upper level vector.}
\label{fig:upperLevel}
\end{minipage}\hfill
\begin{minipage}[t]{0.49\linewidth}
\begin{center}
\epsfig{file=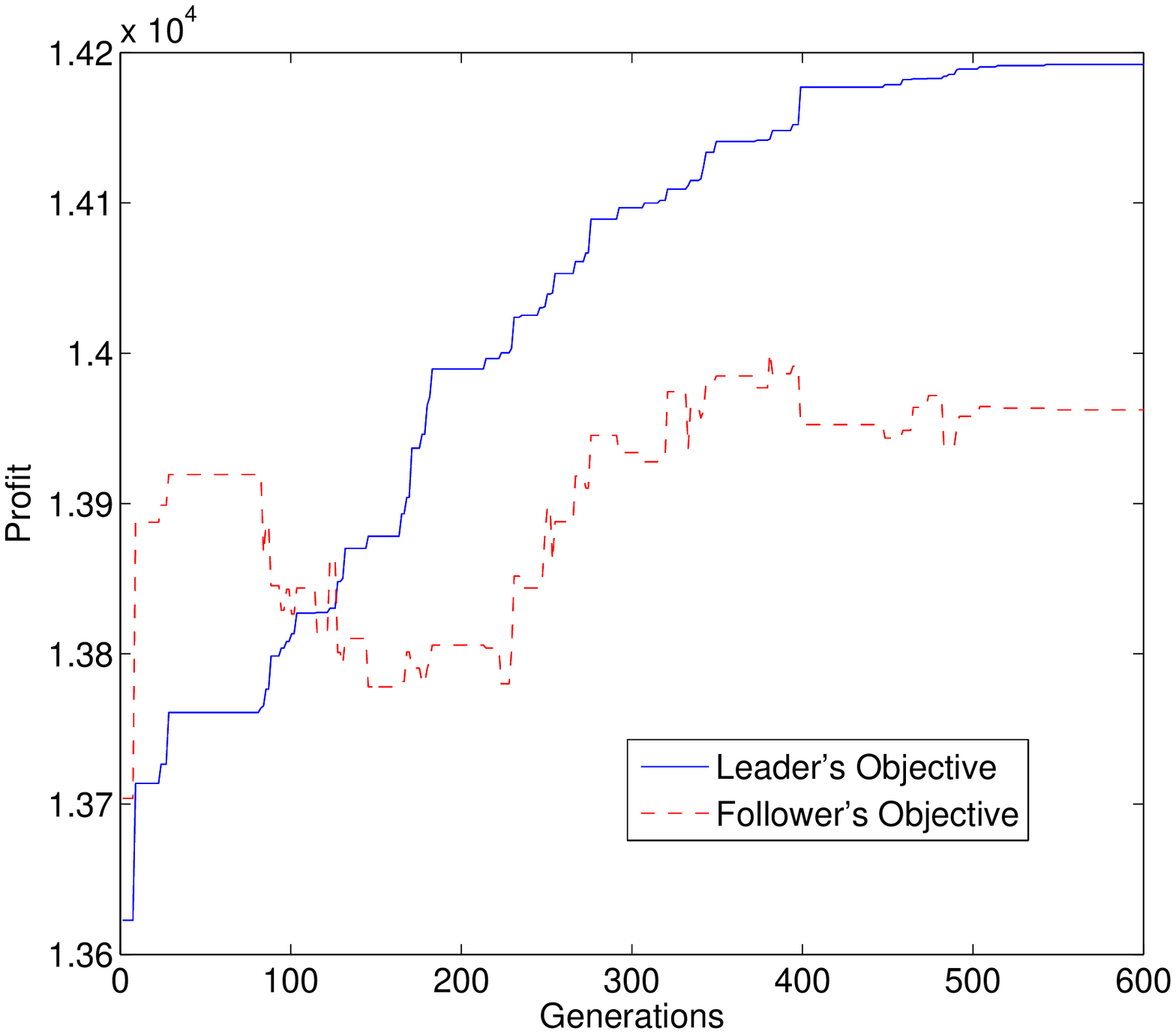,width=\linewidth} 
\end{center}
\caption{Convergence plot showing leader's and follower's objective function over generations.}
\label{fig:convergence}
\end{minipage}
\end{figure*}

The complexity of the problem can be recognized by observing the leader's objective function with respect to the production variables in Figure \ref{fig:upperLevel}, when the follower's decision vectors are at their optimum. The uneven surface at the upper level is due to the discrete nature of the lower level problem. Figure \ref{fig:convergence} provides the convergence plot for the algorithm over generations. The plot shows the leader's objective function value and the follower's objective function value for the elite individual over generations. It can be observed that the leader's objective function value continuously improves, whereas this is not necessary for the follower's objective function value.

\begin{table}
\caption{Optimal leader-follower decisions for a simple 2-period single-leader single-follower problem.}
\begin{center}
    \begin{tabular}{|c|c|c|c|c|c|}
        \hline
Leader & Follower \\ \hline
$q_{l}^{1}=486$  &  $q_{f}^{1}=463$\\
$q_{l}^{2}=585$ & $q_{f}^{2}=573$\\
$I_{l}^{1}=0$  & $I_{f}^{1}=0$\\
$I_{l}^{2}=1.2831e+03$  & $I_{f}^{2}=1.2542e+03$\\
$M_{l}^{1}=0$  & $M_{f}^{1}=0$\\
$M_{l}^{2}=6.4155e+02$  & $M_{f}^{2}=6.2708e+02$\\
$\Psi_{l}=1.4191e+04$ & $\Psi_{f}=1.3963e+04$\\
        \hline
    \end{tabular}
\end{center}
\label{tab:simple}
\vspace{0mm}
\end{table}

\section{Conclusions and Future Work}
\label{sec:conclusion}
In this paper, we have presented a formulation of a multi-period, multi-leader-follower Stackelberg competition model with $N$ leaders, $M$ followers, and $T$ time periods. The formulation can be applied to oligopolistic markets, and one possible application has been briefly discussed. A nested bilevel evolutionary algorithm was used to solve the problem. The ability of the algorithm towards solving difficult bilevel problems has been tested on a recently proposed test-suite of SMD test problems. 

We have solved a concrete version of the multi-period, multi-leader-follower Stackelberg competition model that included $2$ leaders, $5$ followers, and $5$ time periods by employing the nested bilevel evolutionary framework. The approach is computationally intensive, but is easily parallelizable. It allows us to solve the problem despite the difficulties presented by certain assumptions of the model that conventional bilevel methodologies would not be able to overcome. An analysis of the model has also been performed by varying the number of leaders and followers in the system.

Our future efforts would be directed at improving the overall efficiency of the solution methodology so that it can handle more complex problems with a higher number of variables. Additional intelligence needs to be incorporated into the algorithm so that it converges to the optimal solution more quickly. There is also room to improve the problem formulation to reflect the real-world market situations more accurately. A multi-objective extension of the problem would be of further interest.

\section{Acknowledgments}
The authors wish to thank Academy of Finland (Grant: 13133387), Liikesivistysrahasto and HSE Foundation for supporting this study.







\begin{thebibliography}{31}
\providecommand{\natexlab}[1]{#1}
\providecommand{\url}[1]{\texttt{#1}}
\providecommand{\urlprefix}{URL }
\expandafter\ifx\csname urlstyle\endcsname\relax
  \providecommand{\doi}[1]{doi:\discretionary{}{}{}#1}\else
  \providecommand{\doi}{doi:\discretionary{}{}{}\begingroup
  \urlstyle{rm}\Url}\fi
\providecommand{\eprint}[2][]{\url{#2}}
\providecommand{\BIBand}{and}
\providecommand{\bibinfo}[2]{#2}
\ifx\xfnm\undefined \def\xfnm[#1]{\unskip,\space#1}\fi
\bibitem[{Bard(1998)}]{bilevel-book}
\bibinfo{author}{Bard\xfnm[ J.]}.
\newblock \bibinfo{title}{Practical Bilevel Optimization: Algorithms and
  Applications}.
\newblock \bibinfo{publisher}{The Netherlands: Kluwer}; \bibinfo{year}{1998}.
\bibitem[{Stackelberg(1952)}]{Stackelberg52}
\bibinfo{author}{Stackelberg\xfnm[ H.v.]}.
\newblock \bibinfo{title}{{The theory of the market economy.}}
\newblock \bibinfo{address}{New York}: \bibinfo{publisher}{Oxford University
  Press}; \bibinfo{year}{1952}.
\bibitem[{Sherali(1984)}]{Sherali84}
\bibinfo{author}{Sherali\xfnm[ H.D.]}.
\newblock \bibinfo{title}{{A Multiple Leader Stackelberg Model and Analysis}}.
\newblock \bibinfo{journal}{Operations Research} \bibinfo{year}{March/April
  1984};\bibinfo{volume}{32}(\bibinfo{number}{2}):\bibinfo{pages}{390--404}.
\bibitem[{Sherali et~al.(1983)Sherali, Soyster and Murphy}]{Sherali83}
\bibinfo{author}{Sherali\xfnm[ H.D.]}, \bibinfo{author}{Soyster\xfnm[ A.L.]},
  \bibinfo{author}{Murphy\xfnm[ F.H.]}.
\newblock \bibinfo{title}{{Stackelberg-Nash-Cournot Equilibria:
  Characterizations and Computations}}.
\newblock \bibinfo{journal}{Operations Research} \bibinfo{year}{March/April
  1983};\bibinfo{volume}{31}(\bibinfo{number}{2}):\bibinfo{pages}{253--276}.
\bibitem[{Frantsev et~al.(2012)Frantsev, Sinha and Malo}]{cao12}
\bibinfo{author}{Frantsev\xfnm[ A.]}, \bibinfo{author}{Sinha\xfnm[ A.]},
  \bibinfo{author}{Malo\xfnm[ P.]}.
\newblock \bibinfo{title}{{Finding Optimal Strategies in Multi-period
  Stackelberg Games Using an Evolutionary Framework}}.
\newblock In: \bibinfo{booktitle}{15th IFAC Workshop on Control Applications of
  Optimization (CAO'2012)}. \bibinfo{year}{2012},.
\bibitem[{DeMiguel and Xu(2009)}]{DeMiguel09}
\bibinfo{author}{DeMiguel\xfnm[ V.]}, \bibinfo{author}{Xu\xfnm[ H.]}.
\newblock \bibinfo{title}{{A Stochastic Multiple-Leader Stackelberg Model:
  Analysis, Computation, and Application}}.
\newblock \bibinfo{journal}{Operations Research}
  \bibinfo{year}{September/October
  2009};\bibinfo{volume}{57}(\bibinfo{number}{5}):\bibinfo{pages}{1220--1235}.
\bibitem[{Lung and Dumitrescu(2008)}]{lung08}
\bibinfo{author}{Lung\xfnm[ R.I.]}, \bibinfo{author}{Dumitrescu\xfnm[ D.]}.
\newblock \bibinfo{title}{Computing nash equilibria by means of evolutionary
  computation.}
\newblock \bibinfo{journal}{International Journal of Computers, Communications
  and Control}
  \bibinfo{year}{2008};\bibinfo{volume}{3}:\bibinfo{pages}{364--368}.
\bibitem[{Koh(2012)}]{koh12}
\bibinfo{author}{Koh\xfnm[ A.]}.
\newblock \bibinfo{title}{An evolutionary algorithm based on nash dominance for
  equilibrium problems with equilibrium constraints}.
\newblock \bibinfo{journal}{Appl Soft Comput}
  \bibinfo{year}{2012};\bibinfo{volume}{12}(\bibinfo{number}{1}):\bibinfo{pages}{161--173}.
\bibitem[{Lu et~al.(2007)Lu, Shi, Zhang and Ruan}]{lu07}
\bibinfo{author}{Lu\xfnm[ J.]}, \bibinfo{author}{Shi\xfnm[ C.]},
  \bibinfo{author}{Zhang\xfnm[ G.]}, \bibinfo{author}{Ruan\xfnm[ D.]}.
\newblock \bibinfo{title}{An extended branch and bound algorithm for bilevel
  multi-follower decision making in a referential-uncooperative situation}.
\newblock \bibinfo{journal}{International journal of information technology and
  decision-making}
  \bibinfo{year}{2007};\bibinfo{volume}{6}(\bibinfo{number}{2}):\bibinfo{pages}{371--388}.
\bibitem[{Zhang et~al.(2008)Zhang, Lu and Gao}]{zhang08}
\bibinfo{author}{Zhang\xfnm[ G.]}, \bibinfo{author}{Lu\xfnm[ J.]},
  \bibinfo{author}{Gao\xfnm[ Y.]}.
\newblock \bibinfo{title}{Fuzzy bilevel programming: multi-objective and
  multi-follower with shared variables}.
\newblock \bibinfo{journal}{International journal of uncertainty fuzziness and
  knowledge-based systems}
  \bibinfo{year}{2008};\bibinfo{volume}{16}:\bibinfo{pages}{105--133}.
\bibitem[{Colson et~al.(2007)Colson, Marcotte and Savard}]{colson}
\bibinfo{author}{Colson\xfnm[ B.]}, \bibinfo{author}{Marcotte\xfnm[ P.]},
  \bibinfo{author}{Savard\xfnm[ G.]}.
\newblock \bibinfo{title}{An overview of bilevel optimization}.
\newblock \bibinfo{journal}{Annals of Operational Research}
  \bibinfo{year}{2007};\bibinfo{volume}{153}:\bibinfo{pages}{235--256}.
\bibitem[{Vicente and Calamai(2004)}]{vicente-review}
\bibinfo{author}{Vicente\xfnm[ L.N.]}, \bibinfo{author}{Calamai\xfnm[ P.H.]}.
\newblock \bibinfo{title}{Bilevel and multilevel programming: {A} bibliography
  review}.
\newblock \bibinfo{journal}{Journal of Global Optimization}
  \bibinfo{year}{2004};\bibinfo{volume}{5}(\bibinfo{number}{3}):\bibinfo{pages}{291--306}.
\bibitem[{Dempe et~al.(2006)Dempe, Dutta and Lohse}]{dempe-dutta}
\bibinfo{author}{Dempe\xfnm[ S.]}, \bibinfo{author}{Dutta\xfnm[ J.]},
  \bibinfo{author}{Lohse\xfnm[ S.]}.
\newblock \bibinfo{title}{Optimality conditions for bilevel programming
  problems}.
\newblock \bibinfo{journal}{Optimization}
  \bibinfo{year}{2006};\bibinfo{volume}{55}(\bibinfo{number}{5–6}):\bibinfo{pages}{505–--524}.
\bibitem[{Deb and Sinha(2010)}]{my-ecj10}
\bibinfo{author}{Deb\xfnm[ K.]}, \bibinfo{author}{Sinha\xfnm[ A.]}.
\newblock \bibinfo{title}{An efficient and accurate solution methodology for
  bilevel multi-objective programming problems using a hybrid
  evolutionary-local-search algorithm}.
\newblock \bibinfo{journal}{Evolutionary Computation Journal}
  \bibinfo{year}{2010};\bibinfo{volume}{18}(\bibinfo{number}{3}):\bibinfo{pages}{403--449}.
\bibitem[{Bianco et~al.(2009)Bianco, Caramia and Giordani}]{bianco-kkt}
\bibinfo{author}{Bianco\xfnm[ L.]}, \bibinfo{author}{Caramia\xfnm[ M.]},
  \bibinfo{author}{Giordani\xfnm[ S.]}.
\newblock \bibinfo{title}{A bilevel flow model for hazmat transportation
  network design}.
\newblock \bibinfo{journal}{Transportation Research Part C: Emerging
  technologies}
  \bibinfo{year}{2009};\bibinfo{volume}{17}(\bibinfo{number}{2}):\bibinfo{pages}{175--196}.
\bibitem[{Dempe(2002)}]{Dempe02}
\bibinfo{author}{Dempe\xfnm[ S.]}.
\newblock \bibinfo{title}{{Foundations of Bilevel Programming}}.
\newblock \bibinfo{address}{Secaucus, NJ, USA}: \bibinfo{publisher}{Kluwer
  Academic Publishers}; \bibinfo{year}{2002}.
\newblock ISBN \bibinfo{isbn}{9781402006319}.
\bibitem[{Pakala(1993)}]{stackelberg-design}
\bibinfo{author}{Pakala\xfnm[ R.R.]}.
\newblock \bibinfo{title}{A study on applications of stackelberg game
  strategies in concurrent design models}.
\newblock \bibinfo{year}{1993}.
\bibitem[{Herskovits et~al.(2000)Herskovits, Leontiev, Dias and
  Santos}]{bilevel-KKT1}
\bibinfo{author}{Herskovits\xfnm[ J.]}, \bibinfo{author}{Leontiev\xfnm[ A.]},
  \bibinfo{author}{Dias\xfnm[ G.]}, \bibinfo{author}{Santos\xfnm[ G.]}.
\newblock \bibinfo{title}{Contact shape optimization: A bilevel programming
  approach}.
\newblock \bibinfo{journal}{Struct Multidisc Optimization}
  \bibinfo{year}{2000};\bibinfo{volume}{20}:\bibinfo{pages}{214--221}.
\bibitem[{Bard and Falk(1982)}]{bard82}
\bibinfo{author}{Bard\xfnm[ J.]}, \bibinfo{author}{Falk\xfnm[ J.]}.
\newblock \bibinfo{title}{An explicit solution to the multi-level programming
  problem}.
\newblock \bibinfo{journal}{Computers and Operations Research}
  \bibinfo{year}{1982};\bibinfo{volume}{9}:\bibinfo{pages}{77--100}.
\bibitem[{Aiyoshi and Shimizu(1981)}]{aiyoshi81}
\bibinfo{author}{Aiyoshi\xfnm[ E.]}, \bibinfo{author}{Shimizu\xfnm[ K.]}.
\newblock \bibinfo{title}{Hierarchical decentralized systems and its new
  solution by a barrier method}.
\newblock \bibinfo{journal}{IEEE Transactions on Systems, Man, and Cybernetics}
  \bibinfo{year}{1981};\bibinfo{volume}{11}:\bibinfo{pages}{444--449}.
\bibitem[{Yin(2000)}]{yin-bilevel}
\bibinfo{author}{Yin\xfnm[ Y.]}.
\newblock \bibinfo{title}{Genetic algorithm based approach for bilevel
  programming models}.
\newblock \bibinfo{journal}{Journal of Transportation Engineering}
  \bibinfo{year}{2000};\bibinfo{volume}{126}(\bibinfo{number}{2}):\bibinfo{pages}{115--120}.
\bibitem[{Wang et~al.(2008)Wang, Wan, Wang and Lv}]{GA_Wang}
\bibinfo{author}{Wang\xfnm[ G.]}, \bibinfo{author}{Wan\xfnm[ Z.]},
  \bibinfo{author}{Wang\xfnm[ X.]}, \bibinfo{author}{Lv\xfnm[ Y.]}.
\newblock \bibinfo{title}{Genetic algorithm based on simplex method for solving
  linear-quadratic bilevel programming problem}.
\newblock \bibinfo{journal}{Comput Math Appl}
  \bibinfo{year}{2008};\bibinfo{volume}{56}(\bibinfo{number}{10}):\bibinfo{pages}{2550--2555}.
\bibitem[{Sinha and Deb(2009)}]{my-ifac09}
\bibinfo{author}{Sinha\xfnm[ A.]}, \bibinfo{author}{Deb\xfnm[ K.]}.
\newblock \bibinfo{title}{Towards understanding evolutionary bilevel
  multi-objective optimization algorithm}.
\newblock In: \bibinfo{booktitle}{IFAC Workshop on Control Applications of
  Optimization (IFAC-2009)}; vol.~\bibinfo{volume}{7}.
  \bibinfo{publisher}{Elsevier}; \bibinfo{year}{2009},.
\bibitem[{Naoum-Sawaya and Elhedhli(2011)}]{Sawaya11}
\bibinfo{author}{Naoum-Sawaya\xfnm[ J.]}, \bibinfo{author}{Elhedhli\xfnm[ S.]}.
\newblock \bibinfo{title}{{Controlled predatory pricing in a multiperiod
  Stackelberg game: an MPEC approach}}.
\newblock \bibinfo{journal}{Journal of Global Optimization}
  \bibinfo{year}{2011};\bibinfo{volume}{50}:\bibinfo{pages}{345--362}.
\bibitem[{Sinha et~al.(2005)Sinha, Tiwari and Deb}]{my-cec05}
\bibinfo{author}{Sinha\xfnm[ A.]}, \bibinfo{author}{Tiwari\xfnm[ S.]},
  \bibinfo{author}{Deb\xfnm[ K.]}.
\newblock \bibinfo{title}{A population-based, steady-state procedure for
  real-parameter optimization}.
\newblock In: \bibinfo{booktitle}{2005 IEEE Congress on Evolutionary
  Computation (CEC-2005)}. \bibinfo{publisher}{IEEE Press};
  \bibinfo{year}{2005}, p. \bibinfo{pages}{514--521}.
\bibitem[{Sinha et~al.(2006)Sinha, Srinivasan and Deb}]{my-cec06}
\bibinfo{author}{Sinha\xfnm[ A.]}, \bibinfo{author}{Srinivasan\xfnm[ A.]},
  \bibinfo{author}{Deb\xfnm[ K.]}.
\newblock \bibinfo{title}{A population-based, parent centric procedure for
  constrained real-parameter optimization}.
\newblock In: \bibinfo{booktitle}{2006 IEEE Congress on Evolutionary
  Computation (CEC-2006)}. \bibinfo{publisher}{IEEE Press};
  \bibinfo{year}{2006}, p. \bibinfo{pages}{239--245}.
\bibitem[{Deb et~al.(2002)Deb, Anand and Joshi}]{pcx}
\bibinfo{author}{Deb\xfnm[ K.]}, \bibinfo{author}{Anand\xfnm[ A.]},
  \bibinfo{author}{Joshi\xfnm[ D.]}.
\newblock \bibinfo{title}{A computationally efficient evolutionary algorithm
  for real-parameter optimization}.
\newblock \bibinfo{journal}{Evolutionary Computation Journal}
  \bibinfo{year}{2002};\bibinfo{volume}{10}(\bibinfo{number}{4}):\bibinfo{pages}{371--395}.
\bibitem[{Deb(2000)}]{debpenalty}
\bibinfo{author}{Deb\xfnm[ K.]}.
\newblock \bibinfo{title}{An efficient constraint handling method for genetic
  algorithms}.
\newblock \bibinfo{journal}{Computer Methods in Applied Mechanics and
  Engineering}
  \bibinfo{year}{2000};\bibinfo{volume}{186}(\bibinfo{number}{2--4}):\bibinfo{pages}{311--338}.
\bibitem[{Sinha et~al.(2012)Sinha, Malo and Deb}]{my-cec12a}
\bibinfo{author}{Sinha\xfnm[ A.]}, \bibinfo{author}{Malo\xfnm[ P.]},
  \bibinfo{author}{Deb\xfnm[ K.]}.
\newblock \bibinfo{title}{Unconstrained scalable test problems for
  single-objective bilevel optimization}.
\newblock In: \bibinfo{booktitle}{2012 IEEE Congress on Evolutionary
  Computation (CEC-2012)}. \bibinfo{publisher}{IEEE Press};
  \bibinfo{year}{2012},.
\bibitem[{Jun and Vives(2004)}]{Jun04}
\bibinfo{author}{Jun\xfnm[ B.]}, \bibinfo{author}{Vives\xfnm[ X.]}.
\newblock \bibinfo{title}{{Strategic incentives in dynamic duopoly}}.
\newblock \bibinfo{journal}{Journal of Economic Theory}
  \bibinfo{year}{2004};\bibinfo{volume}{116}(\bibinfo{number}{2}):\bibinfo{pages}{249--281}.
\bibitem[{Karray and Mart{\'{i}}n-Herr{\'{a}}n(2009)}]{Karray09}
\bibinfo{author}{Karray\xfnm[ S.]},
  \bibinfo{author}{Mart{\'{i}}n-Herr{\'{a}}n\xfnm[ G.]}.
\newblock \bibinfo{title}{{A dynamic model for advertising and pricing
  competition between national and store brands}}.
\newblock \bibinfo{journal}{European Journal of Operational Research}
  \bibinfo{year}{2009};\bibinfo{volume}{193}(\bibinfo{number}{2}):\bibinfo{pages}{451
  -- 467}.

\end{thebibliography}


\end{document}